\documentclass[journal]{IEEEtran}

\newcommand{\figwidth}{3.2in}

\newcommand{\textmn}[1]{\textit{#1}}
\newcommand{\dd}[2]{\frac{\partial #1}{\partial #2}}
\newcommand{\ddd}[2]{\frac{\partial^2 #1}{\partial #2^2}}
\newcommand{\bra}[1]{\langle #1|}
\newcommand{\ket}[1]{|#1\rangle}
\newcommand{\braket}[2]{\langle #1|#2\rangle}
\usepackage{color}
\newcommand{\RevA}[1]{#1}
\newcommand{\RevB}[1]{#1}
\usepackage{cite}
\usepackage[cmex10]{amsmath}
\usepackage{bm}
\ifCLASSINFOpdf
  \usepackage[pdftex]{graphicx}  
  \graphicspath{{./fig/}{../fig/}{./}}  
  \DeclareGraphicsExtensions{.pdf,.tif,.tiff}
\else 
  \usepackage[dvips]{graphicx}  
  \graphicspath{{./fig/}{../fig/}{./}}  
  \DeclareGraphicsExtensions{.eps,.tif,.tiff}
\fi
\usepackage{hyperref} 
\usepackage{xcolor} 
\hypersetup{
		colorlinks=true,
		linkcolor={blue!50!black},
		anchorcolor={blue!50!black},
		citecolor={blue!50!black},
		urlcolor={blue!50!black},
		pdfborder={0 0 0},
		bookmarksnumbered=true,
}
\newcounter{MYtempeqncnt}
\begin{document}
\title{Extension of Marcatili's analytical approach for rectangular silicon optical waveguides}
\author{Wouter~J.~Westerveld, 
				Suzanne M. Leinders,
				Koen W.A. van Dongen,								
        H.~Paul~Urbach,        
        and~Mirvais~Yousefi 

\thanks{
This article appeared in IEEE/OSA Journal of Lightwave Technology, vol. 30, no. 14, July 15, 2012. This work was supported by TNO, Delft, The Netherlands, and the IOP Photonic Devices Program of NL Agency. }%
\thanks{RECTWG, an open-source Matlab implementation of the methods presented in this article is available at \url{http://waveguide.sourceforge.net}.}%
\thanks{W. J. Westerveld is with the Optics Research Group, Faculty of Applied Sciences, Delft University of Technology, 2628CJ Delft, The Netherlands, and also with TNO, 2628 CK Delft, The Netherlands.}%
\thanks{S.M. Leinders and K. W. A. van Dongen are with the Laboratory of Acoustical Wavefield Imaging, Faculty of Applied Sciences, Delft University of Technology, 2628 CJ Delft, The Netherlands.}%
\thanks{H. P. Urbach is with the Optics Research Group, Faculty of Applied Sciences, Delft University of Technology, 2628 CJ Delft, The Netherlands (e-mail: h.p.urbach@tudelft.nl).}%
\thanks{M. Yousefi is with Photonic Sensing Solutions, 1013 EN Amsterdam, The Netherlands.}%
\thanks{Digital Object Identifier \href{http://dx.doi.org/10.1109/JLT.2012.2199464}{10.1109/JLT.2012.2199464}}}%
\markboth{This article appeared in JOURNAL OF LIGHTWAVE TECHNOLOGY, VOL. 30, NO. 14, JULY 15, 2012 (\href{http://dx.doi.org/10.1109/JLT.2012.2199464}{DOI:10.1109/JLT.2012.2199464})}{Westerveld \MakeLowercase{\textit{et al.}}: Extension of Marcatili's analytical approach for rectangular silicon optical waveguides}%
\maketitle

\begin{abstract}
Marcatili's famous approximate analytical description of light propagating through rectangular dielectric waveguides, published in 1969, gives accurate results for low-index-contrast waveguides. However, photonic integrated circuit technology has advanced to high-index-contrast (HIC) waveguides. 
In this paper, we improve Marcatili's model by adjusting the amplitudes of the components of the electromagnetic fields in his description. 
We find that Marcatili's eigenvalue equation for the propagation constant is also valid for HIC waveguides. 
Our improved method shows much better agreement with rigorous numerical simulations, in particular for the case of HIC waveguides. 
We also derive explicit expressions for the effective group index and the effects of external forces on the propagation constant. \RevA{Furthermore, with our method the phenomenon of avoided crossing of modes is observed and studied.}
\end{abstract}


\IEEEpubid{\begin{minipage}{\textwidth}\ \\[12pt]
\begin{center}
\copyright 2012 IEEE. Personal use of this material is permitted. Permission from IEEE must be obtained for all other uses, in any current or future media, including reprinting/republishing this material for advertising or promotional purposes, creating new collective works, for resale or redistribution to servers or lists, or reuse of any copyrighted component of this work in other works.
\end{center}\end{minipage}} 

\IEEEpubidadjcol

\section{Introduction}
\label{sec:intro}

The propagation of light through rectangular dielectric optical waveguides cannot be described in closed analytical form. Marcatili's famous approximate analytical approach \cite{refs:Marcatili69} has been used since the 1970's and is treated in many textbooks on optical waveguides theory \cite{refs:marcuse, refs:yeh, refs:pollock, refs:hunsperger2002}.  His method is, however, derived for waveguides with a low refractive index contrast, while nowadays technology has advanced to high-index-contrast (HIC) waveguides. 
Silicon-on-insulator (SOI) technology has, for example, become one of the focus platforms for integrated optics over the last decade. The large refractive index contrast of the materials allows for small device footprint. High-yield mass fabrication is provided using CMOS processes from the electronics industry, that have been tailored to photonic applications \cite{refs:dumon08}. Behavior of integrated optical components, such as ring resonator filters or arrayed waveguide grating (AWG) based multiplexers depend critically on the exact knowledge of the propagating modes in the waveguide \cite{refs:yariv00,refs:smit88}. 
Although numerical solutions such as the circular harmonics method, the film mode matching (FMM) method, \RevA{the variational mode expansion method (VMEM),} or the finite element method (FEM) are available \cite{refs:goell69,refs:subdo94,refs:hammer07,refs:fimmwave}, we believe that an analytical model is useful in order to gain insight in the physics of the devices, and also for fast explorative simulations of photonic integrated circuitry \cite{refs:melloni09}.

In this paper, we extend the range of waveguides for which Marcatili's approximate approach can be applied, in particular to high-index-contrast waveguides. 
Similar to Marcatili, we use an \textit{ansatz} for the form of the modal fields that is based on separation of variables in the waveguide core.
The large index contrast causes, in Marcatili's original approach, a severe mismatch of the electromagnetic fields just inside and just outside the core of the waveguide. 
We show that Marcatili's eigenvalue equation for the propagation speed of the light through waveguides is, in fact, more general and we obtain improved modal electromagnetic fields for the same eigenvalue equation, which have a much lower mismatch. An analytical description is presented, and is compared with the fundamental mismatch of this \textit{ansatz}, which is found by means of an optimization algorithm. 

\IEEEpubidadjcol

Next to this, explicit equations are derived for the effective group index and for the linearized influence of external effects on the effective index of the modes. As an example, we analytically calculate the influence of temperature on the effective index of the modes in the waveguide. Also, results are presented on photonic evanescent field sensors, where the refractive index of the medium in the vicinity of the waveguide is probed with the evanescent tail of the waveguide mode \cite{refs:densmore06, refs:devos07}.

Throughout this paper, we test the analysis with the first three modes in a typical SOI waveguide with a guiding layer height of 300 nm, with infrared light that has a free-space wavelength $\lambda_0 = 1550$ nm. These guides consist of a thin monocrystalline silicon layer ($n$ = 3.476) on top of a thick silicon dioxide (BOX) layer ($n$ = 1.444) \cite{refs:fimmwavematerial}. The influence of the silicon substrate below the BOX layer is neglected. 

In the next section, we present our extension of Marcatili's approach. 
In Sec. \ref{sec:novapp}, we apply the eigenvalue equation for the effective index and derive explicit equations for the effective group index and for the effects of external forces. 
In Sec. \ref{sec:numerical}, we compare our theory with rigorous simulations of typical SOI waveguides. Section \ref{sec:conclusion} concludes this paper.

\section{Theory}
\label{sec:theory}

In this section, we describe an approximate model of how light travels through rectangular high-index-contrast (HIC) dielectric waveguides. 
In Sec. \ref{sec:rectwg}, we make an \textit{ansatz} for modes in a rectangular waveguide, and briefly investigate the implications of this \textit{ansatz} at the interfaces of the waveguide core. In Sec. \ref{sec:Marcatili}, we describe Marcatili's choice of parameters for the \textit{ansatz} and Sec. \ref{sec:improved} presents our improved method. In Sec. \ref{sec:lmfit}, we present a quantification of the error that arises from the discontinuities of the electromagnetic field at the interfaces, and we propose an optimization method based on this quantification. In Sec. \ref{sec:lmfitdiss}, we use this quantification to discuss the different methods.

\subsection{Rectangular waveguides} 
\label{sec:rectwg}
\begin{figure}[tb]
\centering
\includegraphics[width=\figwidth]{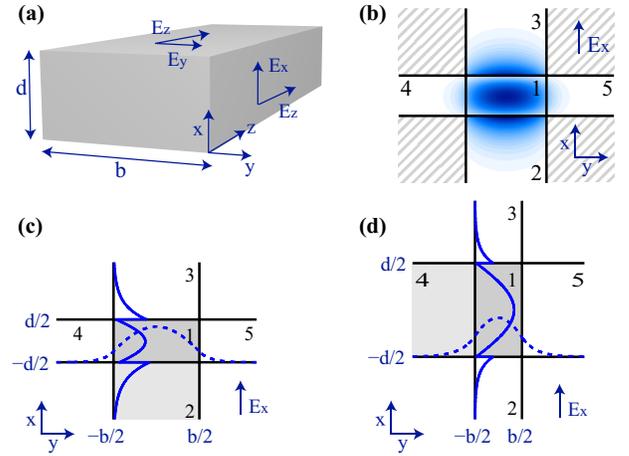} 
\caption{\label{fig:sketch} Schematics of a rectangular waveguide with a width of 600 nm and a height of 300 nm, used to guide light with a free-space wavelength of 1550 nm. (a) 3-D sketch of the waveguide, also displaying the tangential electric field components. (b) Cross-section of the waveguide at $z=0$. Regions~1 to 5 are indicated. The shaded corner regions are neglected in this analysis.  The color plot represents $E_x$, the dominant electric field component, of the fundamental TM-like mode. (c) Sketch of the cross-section of the waveguide with the coordinate frame for a typical TM-like mode. Shape of $E_x$ in the x direction (solid line) and in the y-direction (dashed line). Refractive indices $n_1 = 3.476$, $n_2 = 1.444$, $n_3 = n_4 = n_5 = 1$. (d) Sketch of the cross-section of the waveguide with the coordinate frame for a typical TE-like mode. The waveguide geometry is rotated such that $E_x$ is tangential to the ``upper'' surface of the waveguide. Refractive indices $n_1 = 3.476$, $n_4 = 1.444$, $n_2 = n_3 = n_5 = 1$. The mode profiles in this figure are calculated using the \textmn{amplitude optimization method}.}
\end{figure}

In regular lossless dielectric waveguides, the core of the waveguide has a higher refractive index ($n_1$) than the surrounding media ($n_2$-$n_5$), as depicted in Fig \ref{fig:sketch}. The refractive indices for the outer quadrants, i.e. in the corner regions, are not specified because we neglect these regions. We consider a monochromatic wave with angular frequency $\omega$, propagating in the waveguide direction (z) with a propagation constant $\beta$. For a two-dimensional refractive index profile $n(x,y)$, solutions of Maxwell's equations can be found in the form
\begin{equation} \label{eq:monowave}
\bm{\mathcal{E}}(x,y,z,t) = \textup{Re}\left\{ \bm{E}(x,y) \exp[\imath(\omega t - \beta z)] \right\},
\end{equation} 
and a similar description of $\bm{\mathcal{H}}$, with $\bm{\mathcal{E}}$ and $\bm{\mathcal{H}}$ being the electric and magnetic field, respectively. The free-space propagation constant is $k_0 = \omega / c$, where $c$ is the speed of light in vacuum.
 When the propagation constant $\beta$ is larger than the propagation constant that is allowed in the regions outside the waveguide core, due to spectral cut-off (i.e. $\beta > k_0 n_j$, with $j=2,..,5$), the light is confined in the core of the waveguide. The lateral confinement of such guided waves dictates the light to exist in the form of certain modes, or ``standing waves''.  

Using Maxwell's equations, it is now possible to describe the full electromagnetic fields in terms of the longitudinal field components \cite{refs:marcuse}, i.e. in region~$j$

\begin{eqnarray} 
E_x & = & \frac{-\imath}{K_j^2} \left ( \beta \dd{E_z}{x} + \omega \mu_0 \dd{H_z}{y} \right ), \label{eq:Ex}\\
E_y & = & \frac{-\imath}{K_j^2} \left ( \beta \dd{E_z}{y} - \omega \mu_0 \dd{H_z}{x} \right ), \label{eq:Ey} \\
H_x & = & \frac{-\imath}{K_j^2} \left ( \beta \dd{H_z}{x} - \omega \epsilon_0 n_j^2 \dd{E_z}{y} \right ), \label{eq:Hx}\\
H_y & = & \frac{-\imath}{K_j^2} \left ( \beta \dd{H_z}{y} + \omega \epsilon_0 n_j^2 \dd{E_z}{x} \right ), \label{eq:Hy}
\end{eqnarray}
where $\mu_0$ and $\epsilon_0$ are the permeability and the permittivity of vacuum, and $K_j$ is defined by
\begin{equation}
K_j = \sqrt{n_j^2 k_0^2 - \beta^2}.
\end{equation}
All components, and hence in particular the longitudinal components $E_z$ and $H_z$, satisfy the reduced wave equation (here given for $E_z$) \cite{refs:marcuse}:
\begin{equation}\label{eq:WE}
\ddd{E_z}{x} + \ddd{E_z}{y} + K_j^2 E_z = 0.
\end{equation}
Furthermore, when Eqs. (\ref{eq:Ex})-(\ref{eq:Hy}) are satisfied, the electric and magnetic fields satisfy $\bm{\nabla} \cdot (\epsilon_0 n_j^2 \bm{\mathcal{E}}) = 0$ and $\bm{\nabla} \cdot \bm{\mathcal{H}} = 0$, respectively.

In this section, we adopt a description of the behavior of the light in a rectangular waveguide that is based on separation of spatial variables ($x$, $y$) in the core region, similar to Marcatili's \textit{ansatz}. We neglect the effect of the corners based on the observation that the field is small in those areas. The modal field then consists of standing waves in the core of the waveguide and an exponentially decaying field outside the core (see Fig \ref{fig:sketch}). We will show that the form we propose here can not provide an exact solution of Maxwell's equations, and thus provides an approximate description of the physics. The proposed solution obeys Maxwell's equations in regions~1-5, so that all errors that arise from the approximation show themselves at the interfaces between the waveguide core and its cladding. At these interfaces, continuity of the electromagnetic field components tangential to these interfaces is required (referred to as the electromagnetic boundary conditions), but we allow for discontinuities that are small compared to the field strength in the core of the waveguide. When all tangential components are continuous, the normal components automatically obey Maxwell's equations. 

In our analysis, we propose a form where the electric field is predominantly polarized in the x-direction. From symmetry, the field could as well be chosen predominantly polarized in the y-direction as there is no further discrimination between the x- and the y-direction. Sketches (a)-(c) in Fig.~\ref{fig:sketch} show the fundamental TM-like mode in a waveguide with a width of 600 nm and a height of 300 nm. Sketch (d) shows the fundamental TE-like mode in this waveguide. Instead of rotating the coordinate-frame, the waveguide itself was rotated such that the ``top'' surface of the waveguide is tangential to $E_x$. 

We make the  \textit{ansatz} on the form of the longitudinal components of the  modal electromagnetic field in region~1:
\begin{align}
E_z = & A_1 \sin[k_x(x + \xi)] \cos[k_y(y+\eta)], \label{eq:field1}\\
H_z = & A_2 \cos[k_x(x + \xi)] \sin[k_y(y+\eta)]. \label{eq:field1b}
\end{align}
Then the transversal electromagnetic field components inside region 1 follow from Eqs. (\ref{eq:Ex})-(\ref{eq:Hy}) as
\begin{align}
E_x = & \frac{(\beta A_1 k_x + \omega \mu_0 A_2 k_y)}{\imath K_1^2} \cos[k_x(x + \xi)] \cos[k_y(y+\eta)], \label{eq:Ex2} \\
E_y = & \frac{(-\beta A_1 k_y + \omega \mu_0 A_2 k_x)}{\imath K_1^2} \sin[k_x(x + \xi)] \sin[k_y(y+\eta)], \\
H_x = & \frac{(-\beta A_2 k_x + \omega \epsilon_0 n_1^2 A_1 k_y)}{\imath K_1^2} \sin[k_x(x + \xi)] \sin[k_y(y+\eta)] \label{eq:Hx1},\\
H_y = & \frac{(\beta A_2 k_y + \omega \epsilon_0 n_1^2 A_1 k_x)}{\imath K_1^2} \cos[k_x(x + \xi)] \cos[k_y(y+\eta)].
\end{align}
In region~2, we set
\begin{align}
E_z = & A_3 \exp[\gamma_2(x + d/2)] \cos[k_y(y+\eta)], \label{eq:field2a}\\
H_z = & A_4 \exp[\gamma_2(x + d/2)] \sin[k_y(y+\eta)],
\end{align}
in region~3
\begin{align}
E_z = & A_5 \exp[-\gamma_3(x - d/2)] \cos[k_y(y+\eta)],\\
H_z = & A_6 \exp[-\gamma_3(x - d/2)] \sin[k_y(y+\eta)],
\end{align}
in region~4
\begin{align}
E_z = & A_7 \sin[k_x(x + \xi)] \exp[\gamma_4(y+b/2)],\\
H_z = & A_8 \cos[k_x(x + \xi)] \exp[\gamma_4(y+b/2)],
\end{align}
and in region~5
\begin{align}
E_z = & A_9 \sin[k_x(x + \xi)] \exp[-\gamma_5(y-b/2)],\\
H_z = & A_{10} \cos[k_x(x + \xi)] \exp[-\gamma_5(y-b/2)]. \label{eq:field2}
\end{align}
Here, amplitudes $A_1$ - $A_{10}$, spatial frequencies $k_x$, $k_y$, spatial shifts $\xi$, $\eta$, and exponential decay strengths $\gamma_j > 0$, $j=2,..,5$ are still to be determined. The transversal components are calculated from Eqs (\ref{eq:Ex})-(\ref{eq:Hy}). From the wave equation (\ref{eq:WE}) in region~1, we arrive at
\begin{equation} 
K_1^2 = n_1^2 k_0^2 - \beta^2 = k_x^2 + k_y^2,			\label{eq:WE1} \\
\end{equation}
or, with positive $\beta$, 
\begin{equation}\label{eq:WE1a}
\beta = \sqrt{n_1^2 k_0^2 - k_x^2 - k_y^2}. 
\end{equation}
The effective (refractive) index of the mode in the waveguide is defined by $n_{\textup{eff}} = \beta/k_0$.
From the wave equation (\ref{eq:WE}), together with Eq. (\ref{eq:WE1}), we find
\begin{align} 
\gamma_2^2 = (n_1^2 - n_2^2) k_0^2 - k_x^2, 	\label{eq:WE2} \\
\gamma_3^2 = (n_1^2 - n_3^2) k_0^2 - k_x^2, 	\label{eq:WE3} \\
\gamma_4^2 = (n_1^2 - n_4^2) k_0^2 - k_y^2, 	\label{eq:WE4} \\
\gamma_5^2 = (n_1^2 - n_5^2) k_0^2 - k_y^2. 	\label{eq:WE5}
\end{align}
The sine and cosine dependency on $x$ and $y$, of the transversal electric ($E_z$) and magnetic ($H_z$) field components, was chosen such that the dominant electric field component ($E_x$) is described by a cosine function in both $x$ and $y$ direction. For most common waveguides $n_2 \approx n_3$ and $n_4 \approx n_5$, from which follows that spatial shifts $\xi$ and $\eta$ are small. From \RevA{the observation that the field of the fundamental mode of a waveguide has its highest energy density in the center of the guide}, we expect that the field components with a cosine behavior in both x- and y-directions ($E_x$ and $H_y$) carry the majority the field's energy, next the components with a sine and a cosine dependence in x or y ($E_z$ and $H_z$), and finally the least energy is expected in the field components that have a sinusoidal profile in both x- and y-direction ($E_y$ and $H_x$). These components have antinodes (i.e. high energy density) in the corners of the waveguide core. 

This description has $A_1$ - $A_{10}$, $\xi$, $\eta$, $k_x$ and $k_y$ as free parameters. There are 4 interfaces, with 4 tangential field components to be matched per interface, adding up to 16 equations from the electromagnetic boundary conditions. The tangential field components are depicted in Fig.~\ref{fig:sketch}(a). Field amplitude $A_1$ is used as normalization factor, so we end up with an overdetermined system of only 13 free parameters for 16 equations. 

In summary, we proposed an \textit{ansatz} on the form of the electromagnetic fields of the modes in a rectangular dielectric waveguide, such that Maxwell's equations are obeyed in all regions~1-5. This \textit{ansatz} has 13 free parameters. From continuity of the tangential electromagnetic field components at the interfaces, 16 boundary condition equations follow. In the remainder of Sec. \ref{sec:rectwg}, the requirements that follow from continuity at either the surface normal (Sec. \ref{sec:rectwgBCnorm}) or parallel (Sec. \ref{sec:rectwgBCpar}) to the dominant electric field component, are given.

\subsubsection{Obeying the electromagnetic boundary conditions at the interfaces normal to the dominant electric field component} \label{sec:rectwgBCnorm} 
In this section, we derive the requirements that follow from continuity of the fields at the 1-2 and 1-3 interfaces. The dominant electric field component, $E_x$, is orthogonal to these interfaces, so an infinitely wide, in the y-direction ($b \rightarrow \infty$), rectangle will describe a TM mode in a slab waveguide, see Fig.~\ref{fig:sketch}(c). From all eight electromagnetic boundary conditions at these interfaces, we find
\begin{align} 
A_2 = & \frac{\omega \epsilon_0 n_1^2 k_y}{\beta k_x} A_1, \label{eq:TMA2}\\
A_3 = & A_1 \sin[k_x(\xi-d/2)], \label{eq:TMA3}\\
A_4 = & A_2 \cos[k_x(\xi-d/2)], \label{eq:TMA4}\\
A_5 = & A_1 \sin[k_x(\xi+d/2)], \label{eq:TMA5}\\
A_6 = & A_2 \cos[k_x(\xi+d/2)], \label{eq:TMA6}
\end{align}
together with
\begin{eqnarray} 
\tan[k_x(\xi-d/2)] & = &  -\frac{n_1^2}{n_2^2}\frac{\gamma_2}{k_x}, \label{eq:eveTM1} \\
\tan[k_x(\xi+d/2)] & = &  \frac{n_1^2}{n_3^2}\frac{\gamma_3}{k_x}.  \label{eq:eveTM2}
\end{eqnarray}
Equations (\ref{eq:TMA3})-(\ref{eq:TMA6}) follow directly from the continuity of $E_z$ and $H_z$. The continuity of $E_y$ and $H_y$ is most easily verified by substituting Eqs. (\ref{eq:TMA2})-(\ref{eq:eveTM2}) into the remaining electromagnetic boundary conditions. 
With these field amplitudes $A_2$ - $A_6$, the magnetic field component $H_x$ is zero in regions~1, 2 and 3, as follows from Eq. (\ref{eq:Hx}).

The last two equations, (\ref{eq:eveTM1}) and (\ref{eq:eveTM2}), can be recognized as the eigenvalue equations for a TM mode in a slab waveguide \cite{refs:marcuse, refs:pollock}. These eigenvalue equations thus do not only hold for a slab solution where $\partial/\partial y =0$ and $H_y$, $E_x$ and $E_z$ are the non-zero field components, but also for our \textit{ansatz} where there is a variation in the y-direction.  We eliminate $\xi$ from these two equations, and arrive at the functional
\begin{equation} \label{eq:eveTM}
G(k_x, k_0, n_1, n_2, n_3, d) \equiv \tan[k_x d] - \frac{n_1^2 k_x ( n_3^2 \gamma_2+ n_2^2 \gamma_3)}{n_2^2 n_3^2 k_x^2 - n_1^4 \gamma_2 \gamma_3} = 0.
\end{equation}

\subsubsection{Obeying the electromagnetic boundary conditions at the interface parallel to the dominant electric field component} \label{sec:rectwgBCpar}
When we obey all eight boundary conditions at the 1-4 and 1-5 interface, to which the dominant electric field component, $E_x$, is parallel, we find
\begin{align} 
A_2 = & \frac{\beta k_y}{\omega \mu_0 k_x} A_1, \label{eq:rectTEA2}\\
A_7 = & A_1 \cos[k_y(\eta-b/2)], \label{eq:rectTEA7}\\
A_8 = & A_2 \sin[k_y(\eta-b/2)], \label{eq:rectTEA8}\\
A_9 = & A_1 \cos[k_y(\eta+b/2)], \label{eq:rectTEA9}\\
A_{10} = & A_2 \sin[k_x(\xi+b/2)], \label{eq:rectTEA10}
\end{align}
together with
\begin{eqnarray} 
\tan[k_y(\eta-b/2)] & = & -\gamma_4/k_y, 	 \label{eq:eveTE1}\\
\tan[k_y(\eta+b/2)] & = & \gamma_5/k_y.		 \label{eq:eveTE2}
\end{eqnarray}
Eliminating $\eta$ from latter two equations gives
\begin{equation}  \label{eq:eveTE}
F(k_y,k_0,n_1,n_4,n_5,b) \equiv \tan[k_y b] - \frac{k_y(\gamma_4+\gamma_5)}{k_y^2 - \gamma_4 \gamma_5} = 0.
\end{equation}
Equations (\ref{eq:rectTEA7})-(\ref{eq:rectTEA10}) follow from the continuity of $E_z$ and $H_z$. Continuity of $E_x$ and $H_x$ is checked by substituting Eqs. \mbox{(\ref{eq:rectTEA2})-(\ref{eq:eveTE2})} into the four boundary conditions corresponding to these field components at the two interfaces. It follows from Eq.~(\ref{eq:Ey}) that with these field amplitudes $A_2$, $A_7$ - $A_{10}$, the electric field component $E_y$ is zero in regions~1, 4 and 5. Equations (\ref{eq:eveTE1}) and (\ref{eq:eveTE2}) are identical to the eigenvalue equations for a TE mode in a slab waveguide. 

\subsubsection{Conclusion}

The eigenvalue equations that follow from the analysis of slab waveguides, in which invariance of the field in one direction is assumed, are in fact more general. Identical equations follow from the \textit{ansatz} for $E_z$ and $H_z$, i.e. Eqs. (\ref{eq:field1})-(\ref{eq:field2}), and from imposing the boundary conditions on the horizontal or vertical interfaces of the rectangular waveguide.

Obeying all electromagnetic boundary conditions at the interfaces normal to $E_x$ (Sec. \ref{sec:rectwgBCnorm}) demand a different amplitude coefficient of the magnetic field in the core ($A_2$) then the conditions that follow from the interfaces parallel to $E_x$ (Sec. \ref{sec:rectwgBCpar}). Therefore the \textit{ansatz} has no solutions that excactly obeys the electromagnetic boundary conditions at all interfaces simultaneously. The next sections are devoted to different possibilities for choosing the free parameters such that a low mismatch of the fields at the boundaries is achieved.

\subsubsection{Normalization} Throughout this work, $A_1$ is normalized such that the power flux through waveguide regions~1-5 equals unity, i.e.
\begin{equation} \label{eq:TotalFlux}
P = \frac{1}{2} \textup{Re} \left\{ \;\; \iint \limits_{\textup{regions 1-5}} \left( E_x H_y^* - E_y H_x^*  \right) d\textup{x}d\textup{y} \right\} = 1,
\end{equation}
where the integral runs over the regions~1-5.

\subsection{Marcatili's approach}
\label{sec:Marcatili}
Marcatili has developed a widely used analytical approach for low-index-contrast waveguides \cite{refs:Marcatili69}. For propagating modes in these guides, $k_0 n_j \approx \beta$ because modes are not guided otherwise, so $k_x/k_0 n_j$ and $k_y/k_0 n_j$ are much smaller then unity. Therefore those quantities are neglected in second order. This is often referred to as ``far from cutoff'', while, ``close to core material spectral cutoff'' would be more appropriate. 

In Marcatili's work, $H_x$ is set to zero in all regions. With this requirement, all electromagnetic boundary conditions at the 1-2 and 1-3 interface can be satisfied. This was shown in Sec. \ref{sec:rectwgBCnorm}, where the requirement that $H_x = 0$ in regions~1, 2 and 3 followed from the boundary conditions at these interfaces. 
At the 1-4 and 1-5 interfaces approximations are necessary since not all boundary conditions can be satisfied at these interfaces. $H_x$ is set to zero and continuity follows trivially. 
In the approximate matching of $E_x$ across these interfaces, quantities on the order of $(k_x / k_0 n_j)^2$ are neglected, which is reasonable for low-index-contrast waveguides. For these guides, $\sqrt{\mu_0} H_z$ is larger than $\sqrt{\epsilon_0} E_z$, and is matched across the horizontal interfaces, while $E_z$ is not matched. 
From the requirements above, it is found that the eigenvalue equation of this waveguide is identical to the slab eigenvalue equations (\ref{eq:eveTM1}), (\ref{eq:eveTM2}), (\ref{eq:eveTE1}), and (\ref{eq:eveTE2}). We will refer to this approach as  \textmn{Marcatili's $H_x = 0$ method}. 

Although neglecting terms on the order of $(k_x / k_0 n_j)^2$ is valid for low-index-contrast waveguides, this quantity is even larger than unity outside the core region of high-index-contrast waveguides. This approximation introduces a large mismatch in the continuity of $E_x$, which is the dominant electric field component. 

Similarly, another approximate solution is obtained by setting $E_y = 0$ \cite{refs:Marcatili69, refs:yeh}. With this demand, all boundary conditions at the interface parallel to $E_x$ can be satisfied, and mismatches occur at the 1-2 an 1-3 interfaces. Analogue to the approach above, $E_y$ is trivially matched and $E_z$ is matched, but $H_z$ is not. $H_y$, which is the dominant magnetic field component, is matched while neglecting terms on the order of $(k_x / k_0 n_j)^2$. Although the field amplitudes are different to the case where $H_x = 0$ in all regions, the eigenvalue equations of this waveguide description are identical and thus also given by Eqs. (\ref{eq:eveTM1}), (\ref{eq:eveTM2}), (\ref{eq:eveTE1}), and (\ref{eq:eveTE2}). We will refer to this approach as \textmn{Marcatili's {$E_y = 0$} method}.

\subsection{Improved method} 
\label{sec:improved}
In this section, we present two improvements of Marcatili's methods that give a better description of high-index-contrast waveguides. Two choices for the matching of the boundary conditions are presented, one where the fields at the interfaces normal to the dominant electric field component, $E_x$, are continuous, and one where the fields at the interfaces parallel to $E_x$ are continuous. In Sec. \ref{sec:lmfitdiss}, we show that the latter is more accurate for the cases considered in this paper.

We argue that the dominant boundary conditions for determining $k_x$ and $\xi$ are at the 1-2 and 1-3 interfaces, and therefore Eqs. (\ref{eq:eveTM1}) and (\ref{eq:eveTM2}) are used to determine $k_x$ and $\xi$. Similarly, continuity at the 1-4 and 1-5 interfaces is used to determine $k_y$ and $\eta$ using Eq. (\ref{eq:eveTE1}) and (\ref{eq:eveTE2}). This is supported by the argument that the two different approximations of the modal electromagnetic field as presented by Marcatili both yield these eigenvalue equations. 

\subsubsection{Improved $H_x \approx 0$ method} \RevA{In comparison with Marcatili's approach, we remove the discontinuity of the dominant electric field at the cost of the weak magnetic field component, $H_x$, being not continuous across the 1-4 and 1-5 interfaces.} In this approach, we demand continuity of the tangential electromagnetic fields at the 1-2 and 1-3 interfaces, which are normal to $E_x$. This determines the electromagnetic fields in regions~1, 2 and 3 by amplitudes $A_2$-$A_6$, analogue to \textmn{Marcatili's $H_x = 0$ method}. At the 1-4 and 1-5 interfaces, we choose to match $E_x$ and $H_z$ perfectly. The first is the dominant electric field component. The latter is chosen in favor of matching $E_z$ because an infinitely high ($d \rightarrow \infty$) waveguide is identical to a slab waveguide in which $E_z=0$. These requirements determine the electromagnetic field in regions~4 and 5, in which $H_x$ is not necessary zero. We will refer to this method as the \textmn{Improved $H_x \approx 0$ method}. 

This gives the slab eigenvalue equations (\ref{eq:eveTM}) and (\ref{eq:eveTE}), and field amplitude Eqs. (\ref{eq:TMA2})-(\ref{eq:TMA6}), together with 
\begin{align}
A_7 = & A_1 \left( 1+ \frac{k_0^2(n_1^2-n_4^2)}{\beta^2} \right) \cos[k_y(\eta-b/2)], \\
A_8 = & A_2 \sin[k_y(\eta-b/2)], \\
A_9 = &  A_1 \left( 1+ \frac{k_0^2(n_1^2-n_5^2)}{\beta^2} \right) \cos[k_y(\eta+b/2)], \\
A_{10} = & A_2 \sin[k_y(\eta+b/2)].
\end{align}

\subsubsection{Improved $E_y \approx 0$ method} \label{sec:BC14} 
\RevA{This improvement removes the discontinuity of the dominant magnetic field component $H_y$, that was present in \textmn{Marcatili's $E_y = 0$ method}. }
We match all tangential field components at the 1-4 and 1-5 interface, which are parallel to the dominant electric field component. From Sec. \ref{sec:rectwgBCpar} follows that $E_y = 0$ in regions~1, 4, and 5. At the 1-2 and 1-3 interfaces, $H_y$ is matched because it is the dominant magnetic field component, and $E_z$ is matched in favor of $H_z$ because an infinitely wide waveguide ($b \rightarrow \infty$) is identical to a slab waveguide in which $H_z=0$. 

We find the slab eigenvalue equations (\ref{eq:eveTM}) and (\ref{eq:eveTE}), and field amplitude Eqs. (\ref{eq:rectTEA2})-(\ref{eq:rectTEA10}), together with
\begin{align}
A_3 = & A_1 \sin[k_x(\xi-d/2)], \\
A_4 = & A_2 \left( 1 + \frac{k_0^2(n_1^2-n_2^2)}{\beta^2} \right) \cos[k_x(\xi-d/2)], \\
A_5 = & A_1 \sin[k_x(\xi+d/2)], \\
A_6 = & A_2 \left( 1 + \frac{k_0^2(n_1^2-n_3^2)}{\beta^2} \right) \cos[k_x(\xi+d/2)].
\end{align}

\subsection{Least-discontinuity optimization of the \textit{ansatz} parameters}
\label{sec:lmfit}
In Sec. \ref{sec:rectwg}, we presented an \textit{ansatz} on the form of the electromagnetic field for modes in a rectangular waveguide. This \textit{ansatz} was chosen such that Maxwell's equations are satisfied in regions~1-5, so that all errors manifest themselves at the four interfaces of the waveguide core. This error, which is the discontinuity of the tangential electromagnetic field components at these interfaces, is referred to as the mismatch. The measure we adopt to quantify this mismatch, or error, is the average energy density that is associated with these discontinuities. In Sec. \ref{sec:numerical}, we show that this intuitive quantity excellently agrees with rigorous numerical results. This analysis is performed in a cross-section of the waveguide at $z=0$, at time $t=0$, as further longitudinal and temporal behavior follows trivially from Eq. (\ref{eq:monowave}).  We define:
\begin{align}\label{eq:BCmis}
U_{\textup{mm}} & = \frac{\epsilon_0}{4l}  \oint (n^+ + n^-)^2 \cdot \left| \bm{\hat{\nu}} \times \left( \bm{E}^+ - \bm{E}^- \right) \right|^2 d\textup{l} \\
& + \frac{\mu_0}{l} \oint \left| \bm{\hat{\nu}} \times \left(\bm{H}^+ - \bm{H}^- \right)\right|^2 d\textup{l} \nonumber.
\end{align}
The four interfaces of the waveguide are simultaneously described by the integral. The line integral runs along the circumference of the waveguide in the (x,y)-plane, and $l = 2(b+d)$ is the length of this circumference. $\bm{E}^+$ and $\bm{E}^-$ are the electric fields just outside and inside the waveguide core region~1, so that $(\bm{E}^+ - \bm{E}^-)$ represents the discontinuity of this field, and $\bm{\hat{\nu}}$ is a unit vector orthogonal to the waveguide surface. The cross product of $\bm{\hat{\nu}}$ with the discontinuity in the field just selects the tangential components. $n^+$ and $n^-$ are the refractive indices just outside and inside the waveguide. At the interface, an average refractive index $(n^+$+$n^-)/2$ is assumed to calculate the energy density of the electric field components. 

Although $U_{\textup{mm}}$ can be intuitively interpreted as an energy density, we cannot attach a rigorous physical meaning to this quantity. The mismatch in the fields only occurs at interfaces, which have no physical volume. Therefore the energy density cannot be integrated over volume in order to obtain a total energy.

In this section, we use this quantity to propose a new \mbox{(semi-)} analytical method. In the next section, we use this quantity to investigate the methods of the two previous sections (\ref{sec:Marcatili} and \ref{sec:improved}) without resorting to numerical simulations. 

In what we call the \textmn{full optimization method}, the mismatch in the fields at the boundaries, as given by Eq. (\ref{eq:BCmis}), is minimized using an unconstrained nonlinear optimization as implemented in \textsc{Matlab} \cite{refs:matlab}. The initial parameters are calculated from the improved method as presented in Sec.~\ref{sec:BC14}. 

\RevA{The \textmn{full optimization method} may be compared with the variational mode expansion method (VMEM) \cite{refs:hammer07}, which is also based on a separation of variables. However, the methods differ substantially because the modal field \textit{ansatz} differs and also the object function used in the variational principle of VMEM differs from the object function that is minimized in our approach. The VMEM is applicable to more general structures.}

\begin{figure}[tbhp]
\centering
\includegraphics[width=\figwidth]{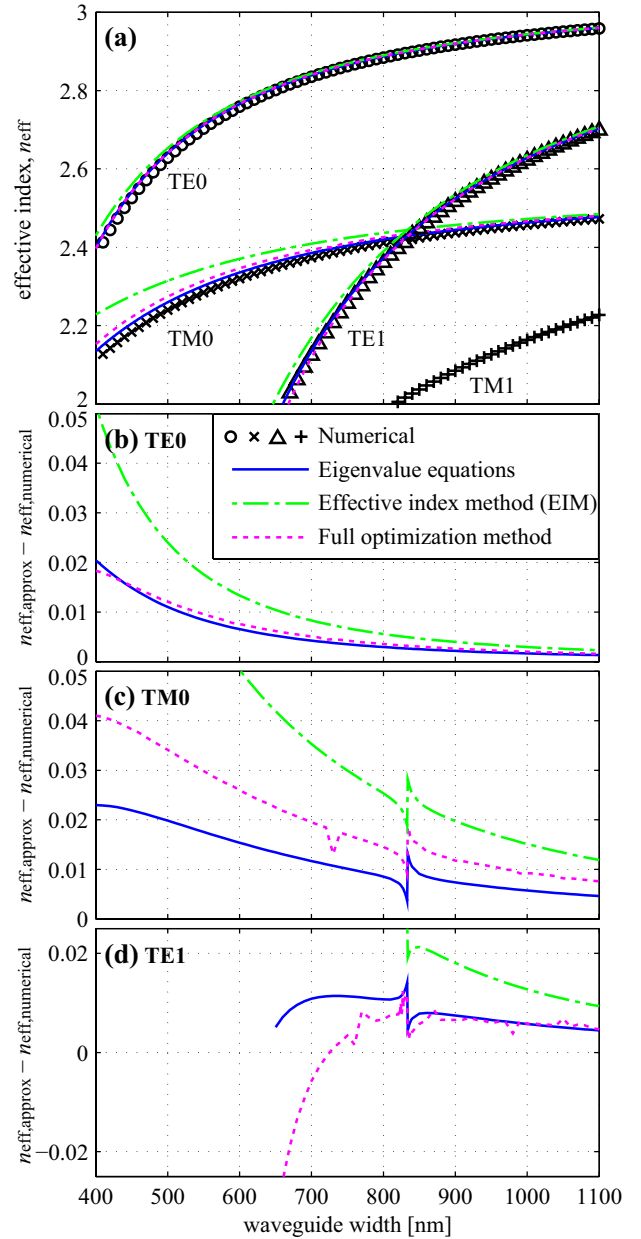} 
\caption{\label{fig:neff} Effective refractive indices calculated using four different methods. Plot (a) presents the first 3 modes in the waveguide core (TE0, TM0, TE1). 
In comparison with conventional notation, e.g. TE$_{00}$, we dropped a zero, as in our waveguide geometries all higher-order modes have higher-order standing waves only in the direction of the width of the waveguide. The numerically calculated effective index of the TM1 mode is included for completeness. Plots (b)-(d) show the difference in the effective index as calculated by the analytical method with respect to the numerical method. }
\end{figure}

\begin{figure}[tbhp]
\centering
\includegraphics[width=\figwidth]{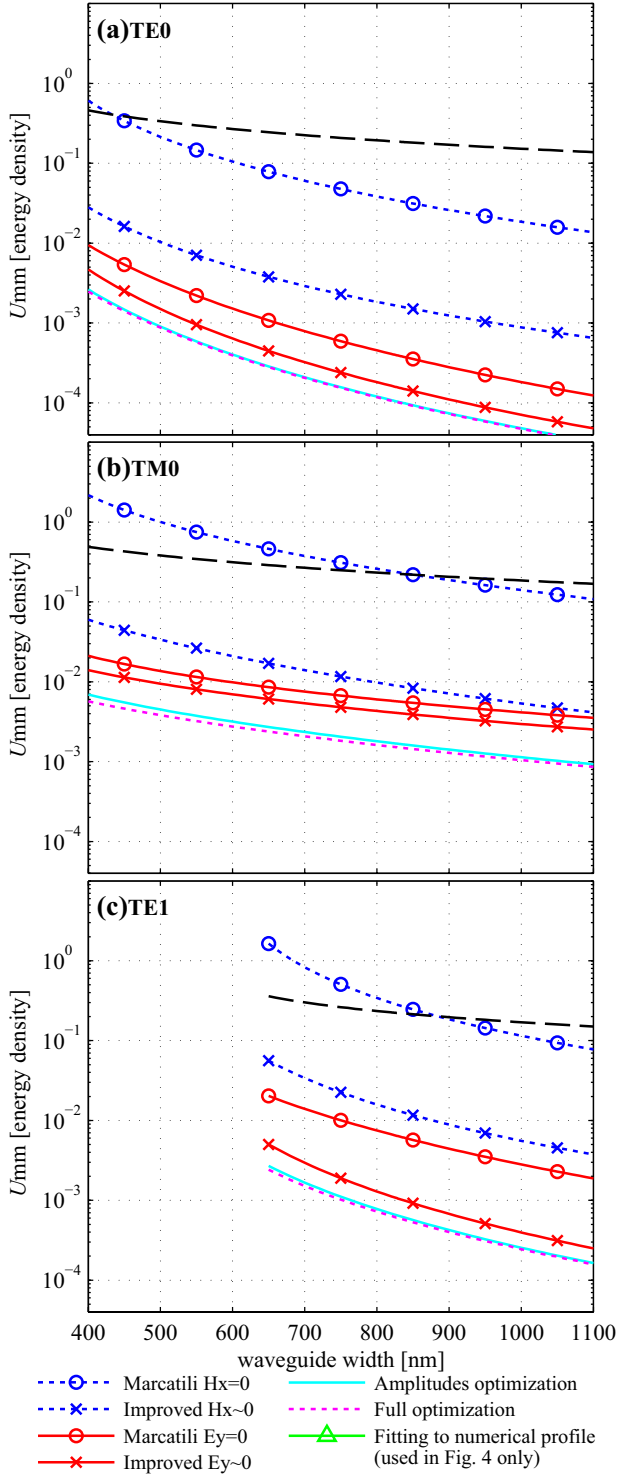}
\caption{\label{fig:mismatch} Mismatch of the tangential field components at the interfaces of the waveguide core, $U_{\textup{mm}}$, see Eq. (\ref{eq:BCmis}). Typical, 300 nm high, SOI waveguide. Plots (a)-(c) represent the TE0, TM0, and TE1 modes in the waveguide. In each plot, 6 different methods to calculate the free parameter in the analytical description of the waveguide mode profile are presented. The black dashed line indicates the average energy density in the core (calculated using the \textmn{full optimization method}). }
\end{figure}

It is interesting to know how accurate $k_x$ and $k_y$, and thus the effective refractive index, are calculated using the analytical methods. The effective indices that follow from the eigenvalue equations on the one hand, and from the least-mismatch optimization on the other hand, are presented in  Fig. \ref{fig:neff}. It can be seen that the difference between the two methods is small. The influence of this difference in propagation speed on the mismatch of the fields at the interfaces is investigated by optimizing the field amplitudes $A_2$ - $A_{10}$ with fixed $k_x$, $k_y$, $\xi$, and $\eta$ as calculated from the analytical eigenvalue equations. As Eq. (\ref{eq:BCmis}) is quadratic in the amplitudes $A_2$ - $A_{10}$, this minimum can be found analytically. This method is referred to as the \textmn{amplitudes optimization method}. As can be seen in Fig.~\ref{fig:mismatch}, the mismatch of the method with the fitted $k$'s ($k_x$ and $k_y$) and method with the analytical calculated $k$'s is almost identical. Therefore we conclude that, with the \textit{ansatz} for the fields as described in Sec. \ref{sec:rectwg}, and the error of the model described by the field mismatch, Eq. (\ref{eq:BCmis}), the values of $k_x$ and $k_y$ are very accurately calculated from analytical slab eigenvalue equations, for the typical SOI waveguides as considered in this paper. 

\subsection{Discussion of the different methods}
\label{sec:lmfitdiss}

In Fig.~\ref{fig:mismatch}, all six different methods are compared. The energy associated with mismatch of the electromagnetic fields at the boundaries of the core of the waveguide, $U_{\textup{mm}}$, is plotted for three types of waveguide modes. \RevA{$U_{\textup{mm}}$ can be interpreted as the energy density of the error at the interfaces. In order to get a feeling for the magnitude of $U_{\textup{mm}}$, the average energy density in the core region of the waveguide is also plotted in this figure (black dashed line).}

For all geometries, the improved methods are better than Marcatili's original methods. Considering the fundamental modes in a 400 nm wide waveguide, and the 1st order TE-like mode in a 650 nm wide waveguide, we see that the mismatch in \textmn{Marcatili's $E_y=0$ method} is 1.5 to 4 times larger than the mismatch in the \textmn{improved $E_y \approx 0$ method}. In Marcatili's method, the power flux through regions~2, 3, 4, or 5, i.e. in the evanescent tail of the mode, is sometimes negative (or in backwards propagating direction). This is an indication of the inaccuracy of the method, and this behavior was not observed for the other methods.

For the waveguide geometries as considered in this paper, the \textmn{improved $E_y \approx 0$ method} has smaller errors in the continuity of the fields at the boundaries than the \textmn{improved $H_x \approx 0$ method}. However, this is geometry dependent, as is expected from the trend in the TM-like mode. For TM-like modes in waveguides wider than 1765 nm ($b > 6d$), the \textmn{improved $H_x \approx 0$ method} has a lower mismatch than the \textmn{improved $E_y \approx 0$ method}. This behavior of the asymptotes is expected from theory of slab waveguides. In the limit $b \rightarrow \infty$, the waveguide is a slab waveguide with interfaces 1-2 and 1-3, so the \textmn{improved $H_x \approx 0$ method} is expected to give better results, because the fields described by this methods are continuous across these interfaces. In the limit $d \rightarrow \infty$, only the 1-4 and 1-5 interfaces play a role, so we expect the \textmn{improved $E_y \approx 0$ method} to give the most accurate results. For the cases considered in this plot, average energy density associated with the mismatch ($U_{\textup{mm}}$) of the \textmn{improved $E_y \approx 0$ method} is below 3\% of the average energy density in the core of the waveguide. 

In conclusion, the effective index of the mode is accurately calculated by the eigenvalue equations. The amplitudes $A_2$ to $A_{10}$ are calculated more accurately with the improved methods than with Marcatili's original methods. A better quantitative evaluation can be made by a comparison of the methods with rigorous numerical calculated modal profiles, which is presented in Sec. \ref{sec:numerical}.

\section{Novel applications of the eigenvalue equation of the propagation constant}
\label{sec:novapp}
In this section, we derive explicit expressions for the effective group index and for the influence of a change in the waveguide.

\subsection{Modal dispersion}
\label{sec:modaldispersion}
The modal dispersion can be described by the (effective refractive) group index:
\begin{equation}
\label{eq:groupindex}
 n_g \equiv n_{\textup{eff}} - \lambda_0 \frac{\partial n_{\textup{eff}}}{\partial \lambda_0} = \frac{ \partial \beta}{\partial k_0}.
\end{equation}
The group index is often used to design photonic integrated circuits. For example to calculate of the free spectral range (FSR) of ring resonators. From Eq. (\ref{eq:WE1a}), we find
\begin{equation}
\label{eq:rectgroupindex}
\dd{\beta}{k_0} = \frac{1}{\beta} \left(k_0 n_1^2 + k_0^2 n_1 \dd{n_1}{k_0} - k_x \dd{k_x}{k_0} - k_y \dd{k_y}{k_0} \right).
\end{equation}
The 1st and 2nd term on the right-hand-side of this equation are specified by the material refractive indices. The refractive indices $n_j(k_0)$ may depend on frequency and thus on $k_0 = \omega/c$. The 3rd term is calculated from Eq. (\ref{eq:eveTM}). Although $k_x$ is only given as implicitly, $\partial k_x / \partial k_0$ can be calculated explicitly. The total derivative of the left-hand-side of Eq. (\ref{eq:eveTM}) with respect to $k_0$, $dG/dk_0$, equals zero for solutions of $G=0$. The height $d$ does not depend on frequency. So we get
\begin{equation}
\label{eq:dGdk0} \nonumber
\frac{dG}{dk_0} = \frac{\partial G}{\partial k_0} 
+ \frac{\partial G}{\partial k_x} \frac{\partial k_x}{\partial k_0}
+ \frac{\partial G}{\partial n_1} \frac{\partial n_1}{\partial k_0}
+ \frac{\partial G}{\partial n_2} \frac{\partial n_2}{\partial k_0}
+ \frac{\partial G}{\partial n_3} \frac{\partial n_3}{\partial k_0}, 
\end{equation}
or, 
\begin{equation}
\label{eq:dkxdk0}
\frac{\partial k_x}{\partial k_0} = -
\frac{
\frac{\partial G}{\partial k_0} 
+ \frac{\partial G}{\partial n_1} \frac{\partial n_1}{\partial k_0}
+ \frac{\partial G}{\partial n_2} \frac{\partial n_2}{\partial k_0}
+ \frac{\partial G}{\partial n_3} \frac{\partial n_3}{\partial k_0}
}
{
\frac{\partial G}{\partial k_x} 
}.
\end{equation}
Similarly, the 4th term of the right-hand-side of Eq. (\ref{eq:rectgroupindex}) is calculated from Eq. (\ref{eq:eveTE}) as
\begin{equation}
\label{eq:dkydk0}
\frac{\partial k_y}{\partial k_0} = -
\frac{
\frac{\partial F}{\partial k_0} 
+ \frac{\partial F}{\partial n_1} \frac{\partial n_1}{\partial k_0}
+ \frac{\partial F}{\partial n_4} \frac{\partial n_4}{\partial k_0}
+ \frac{\partial F}{\partial n_5} \frac{\partial n_5}{\partial k_0}
}
{
\frac{\partial F}{\partial k_y} 
}.
\end{equation}
The partial derivatives in Eqs. (\ref{eq:dkxdk0}) and (\ref{eq:dkydk0}) are straightforward to calculate, as shown in appendix \ref{sec:Amodaldispersion}.

\subsection{Influence of a linear external effect}
\label{sec:rectexternaleffect}
Analogue to the derivation of the effective group index, it is also possible to calculate the linearized influence of an external effect on the propagation constant of the mode. For example, the influence of a temperature change in the waveguide or the influence of a change in the refractive index of the surrounding medium can be calculated. We describe the external effect by a parameter $\chi$. In this theory, the waveguide properties $n_j$, $b$ and $d$ are assumed to be known, as well as the first-order influence of the external effect on these properties, i.e. $\partial n_j / \partial \chi$, $\partial d / \partial \chi$ and $\partial b / \partial \chi$. Taking the derivative of Eq. (\ref{eq:WE1a}) with respect to $\chi$ gives
\begin{equation}
\label{eq:dbetadX}
\dd{\beta}{\chi} = \frac{1}{\beta} \left( n_1 k_0^2 \dd{n_1}{\chi} - k_x \dd{k_x}{\chi} - k_y \dd{k_y}{\chi} \right).
\end{equation}
As in the previous section, $\partial k_x / \partial \chi$ is calculated by taking the derivative of Eq. (\ref{eq:eveTM}), $G = 0$, and solving $\partial k_x / \partial \chi$:
\begin{equation}
\label{eq:dkxdX}
\frac{\partial k_x}{\partial \chi} = -
\frac{
\dd{G}{n_1} \dd{n_1}{\chi}
+ \dd{G}{n_2} \dd{n_2}{\chi}
+ \dd{G}{n_3} \dd{n_3}{\chi}
+ \dd{G}{d} \dd{d}{\chi}
}
{
\frac{\partial G}{\partial k_x} 
}.
\end{equation}
And for $\partial k_y / \partial \chi$, we find from Eq. (\ref{eq:eveTE}) that
\begin{equation}
\label{eq:dkydX}
\frac{\partial k_y}{\partial \chi} = -
\frac{
\dd{F}{n_1} \dd{n_1}{\chi}
+ \dd{F}{n_4} \dd{n_4}{\chi}
+ \dd{F}{n_5} \dd{n_5}{\chi}
+ \dd{F}{b} \dd{b}{\chi}
}
{
\frac{\partial F}{\partial k_y} 
}.
\end{equation}
Equations (\ref{eq:dkxdX}) and (\ref{eq:dkydX}) are given explicitly in Appendix \ref{sec:Aexternaleffect}.

\subsubsection{Influence of temperature on SOI waveguides}
In the case of a temperature change, $\partial n_j / \partial \chi$ is given by the thermo-optic effect. The change in cross-section of the waveguide, $\partial d / \partial \chi$ and $\partial b / \partial \chi$ are described by linear thermal expansion. The thermo-optic coefficients of silicon and silicon dioxide, as well as the linear thermal expansion coefficients, are reported in Refs. \cite{refs:fimmwavematerial} and \cite{refs:okada84}, respectively.

\subsubsection{Evanescent field sensor}
\label{sec:externalcladding}
In evanescent field sensors, liquids with varying refractive indices flow over the SOI waveguides. In this case, $\chi$ describes the liquid refractive index, so that for TM-like modes $n_3 = n_4 = n_5 = \chi$ and $\partial n_3 / \partial \chi = \partial n_4 / \partial \chi = \partial n_5 / \partial \chi = 1$, while the other waveguide properties are constant. For TE-like modes, $n_2 = n_3 = n_5 = \chi$ and $\partial n_2 / \partial \chi = \partial n_3 / \partial \chi = \partial n_5 / \partial \chi = 1$ (see Fig.~\ref{fig:sketch}).

\section{Comparison with rigorous numerical methods}
\label{sec:numerical}
In this section, we compare the analytical results of the previous section with rigorous numerical results. To find the transverse modes of the waveguide, we use the film mode matching (FMM) method as implemented in the FimmWave software package \cite{refs:subdo94, refs:fimmwave}. In this method the cross-section of the ridge waveguide is split in vertical slices, and 1-dimensional modes are computed analytically for each slice. The 2-D modes are now found by simultaneously solving the modal amplitude factors of the 1-D modes in all slices such that they construct a field obeying Maxwell's equations. The area of the numerical simulation extends 2 $\mu$m from the waveguide, and 200 1-D modes are used per slice. \RevA{The width of the waveguide is varied in steps of 10~nm, and in steps of 1~nm between 820~nm and 840~nm.} For verification, we compared the FMM method with the finite element method (FEM) that is also implemented in the FimmWave software, using $\sim$210 gridpoints in both the x- and the y-direction. \RevA{For all datapoints presented in this work, the difference (between FMM and FEM) in effective index is below $10^{-3}$ and the relative energy in the difference field is below $10^{-4}$}. 

The measure that is used to compare two electromagnetic fields is the relative energy in the difference field, i.e.
\begin{equation}\label{eq:FieldsDiff}
\Delta U = \frac{\iint \limits_{\textup{regions 1-5}} 
\left( n^2 \epsilon_0 | \bm{E}^A - \bm{E}^N |^2 
 + \mu_0 |\bm{H}^A - \bm{H}^N|^2 \right) d\textup{x} d\textup{y}}
 {\iint \limits_{\textup{regions 1-5}} \left( n^2 \epsilon_0 |\bm{E}^N|^2 
 + \mu_0 |\bm{H}^N|^2 \right) d\textup{x} d\textup{y}}
\end{equation}
where $\bm{E}^A$ and $\bm{E}^N$ are the analytically and rigorous numerically calculated fields, respectively. This integral runs over all regions that are described by the analytical solution. The integral is calculated analytically with the assumption that the numerical field is piecewise-constant in space.

\begin{figure}[tbhp]
\centering
\includegraphics[width=\figwidth]{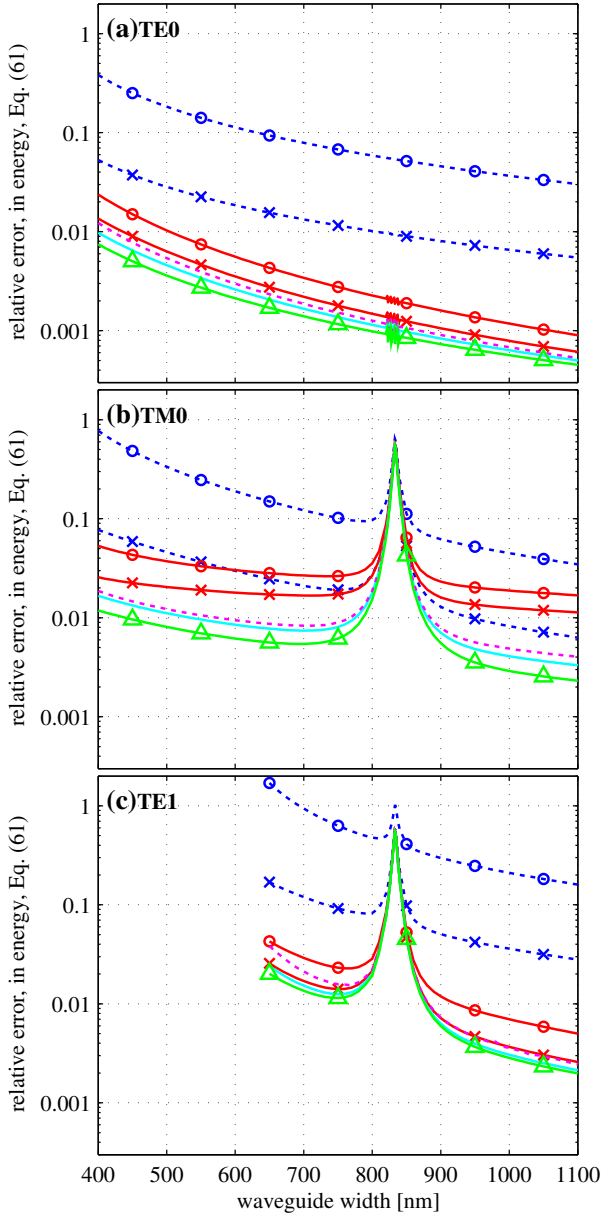} 
\caption{\label{fig:udiff} Energy in the electromagnetic difference field between the analytically calculated fields and the rigorous numerically calculated fields, normalized to the energy in the numerically calculated field. The fields are only considered in regions~1-5. Legend and layout of the figure is identical to Fig.~\ref{fig:mismatch}.}
\end{figure}

\subsection{Improved analytical methods}
The effective refractive indices are calculated using four different methods, including the FMM method, and are presented in Fig.~\ref{fig:neff}. For completeness, results of the effective index method (EIM) \cite{refs:marcuse} are also included. The analytical eigenvalue equations, described in Sec. \ref{sec:rectwg}, give, for most geometries, the most accurate effective index.  The method where the effective index is calculated using least-mismatch fitting in the boundary conditions as presented in Sec. \ref{sec:lmfit} is less accurate for TM-like modes. The error in the EIM becomes significantly large for strong confined modes. 
\RevA{The discontinuities in plots (c) and (d) at a width of 833~nm are discussed in Sec. \ref{sec:crossing}. The other small ``wiggles'' in the \textmn{full optimization method} are due to numerical instabilities. }
It can be seen that the error in the effective index calculated using the analytical eigenvalue equation remains below 2\% for all cases. The error is, for all methods, lower for the weaker-confined modes.

Figure \ref{fig:udiff} presents the energy of the difference field between the analytical and the numerical FMM method. First of all, it is clear that something special happens for the fundamental TM-like (TM0) and the 1st  TE-like (TE1) modes in the waveguide with a width of $\sim$830 nm. We will explain this \RevA{in Sec. \ref{sec:crossing}. The very small ``wiggles'' in the fundamental TE-like mode between waveguide widths of 820~nm and 840~nm are due to our numerical discretization of the electromagnetic fields, which not exactly match the refractive index distribution for features below 10~nm. }

The fundamental limit of the \textit{ansatz} is found by optimizing all free parameters such that the energy in the difference field compared to the numerical result (the quantity plotted in this figure) is minimal.

The method that gives the least difference with the rigorous numerical result is the method where $k_x$, $k_y$, $\xi$, and $\eta$ follow from the slab eigenvalue equations, while the field amplitudes $A_2$-$A_{10}$ are optimized by minimizing the mismatch, $U_{\textup{mm}}$, in Eq. (\ref{eq:BCmis}) (\textmn{amplitude optimization method}). For this method, the energy in the difference field is below 2\% for all waveguide modes considered here (except for the crossing at 830 nm). Also, this method is very close to the fundamental limit of the \textit{ansatz}. 

The fact that the eigenvalue equations work so well, justifies the assumption that the boundary conditions at the 1-2 and 1-3 interfaces can be best used to calculate $k_x$ and the boundary conditions on the 1-4 and 1-5 interfaces can be best used to calculate $k_y$. Deviating from these $k$-values to minimize the total mismatch in the boundary conditions only slightly reduces this mismatch and increases the error in the $k$-values. Still, the quantification of the mismatch as done in Sec. \ref{sec:lmfit} gives surprisingly accurate results. The difference between the fundamental limit of the \textit{ansatz} and both the \textmn{amplitude optimization method} and the \textmn{full optimization method} is very small, especially considering that there is a fundamental error in the \textit{ansatz}, and that the error of the fields only manifests itself at the interfaces of the waveguide core, which is difficult to interpret from a physical perspective.

The improved approaches are, for all cases, more accurate then Marcatili's original approaches. For these methods, the full field-profiles are easily and intuitively calculated. The fields of the SOI waveguides considered in this work are accurately described by the \textmn{improved $E_y \approx 0$ method}, with less than 3\% energy in the difference field (except for the crossing at 830 nm). 

The \textmn{improved $E_y \approx 0$ method} is more accurate than the \textmn{improved $H_x \approx 0$ method} for ($b < 3d$), i.e. for TE-like modes and for TM-like modes in waveguides that are not very wide. From the mismatch in the boundary conditions, the same trend was expected and this crossing was expected at ($b \approx 6 d$). Also, the difference between the \textmn{improved $E_y \approx 0$ method}, the \textmn{improved $H_x \approx 0$ method} and the optimization methods follows the same trend as expected from Sec. \ref{sec:lmfit}.

In conclusion, the improved Marcatili's method describes the effective indices and the field distributions of typical SOI waveguides sufficiently accurately for many applications, \RevA{except when the effective indices of two modes are similar. In the next section, the modes found from our approximation are used to explain what happens in these regions}. When compared to rigorous numerical solutions, the error in the effective index was below 2\% and the relative error in terms of the energy of the fields was less than 3\%. In addition, the optimization method gives even more accurate results for the fields.

\begin{figure}[bt]
\centering
\includegraphics[width=\figwidth]{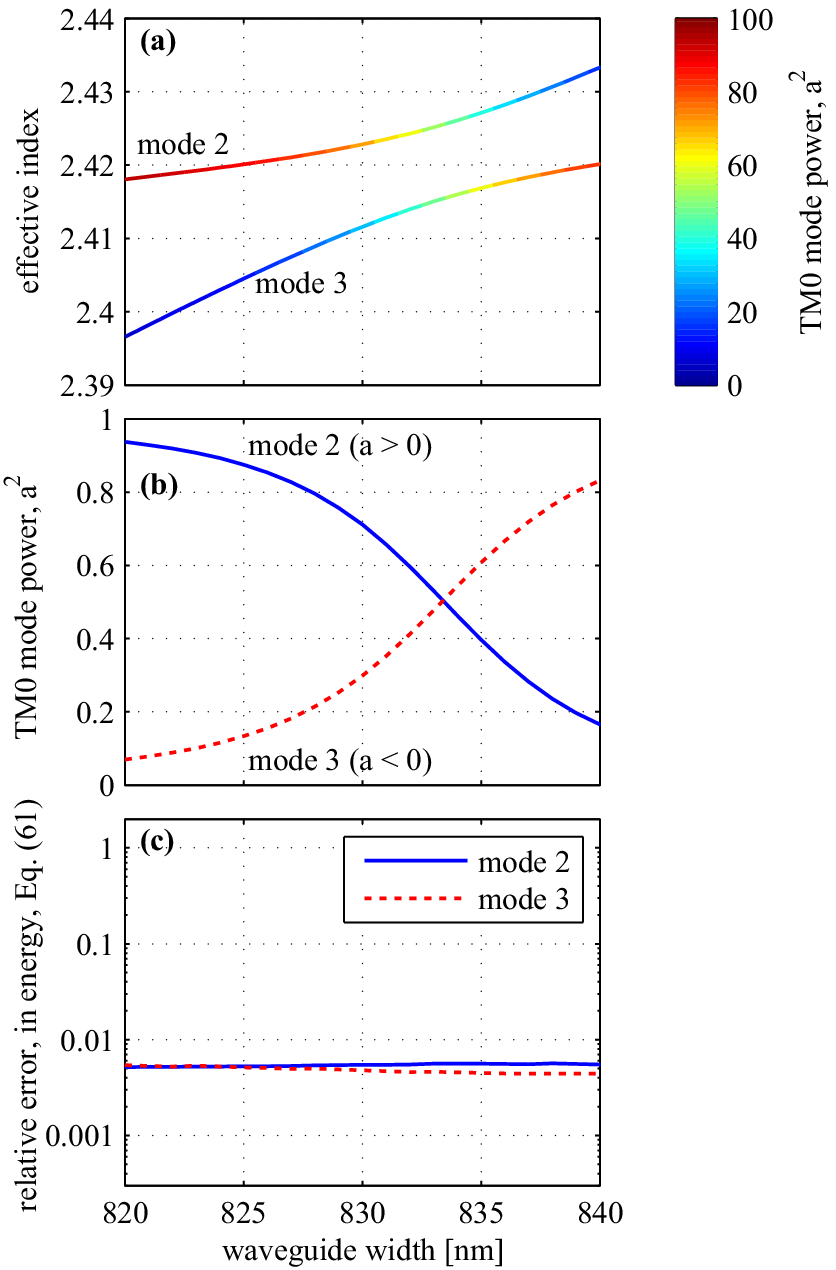} 
\caption{\label{fig:crossing} \RevA{Investigation of the avoided crossing of effective indices of two modes. Plot (a): numerically calculated effective indices of the 2nd and 3rd mode, zoom-in of Fig.~\ref{fig:neff}(a). Plot (b): power in the TM0-like mode when the fields of the modes in plot (a) are written as a superposition of a TM0-like and a TM1-like mode. The curves in plot (a) are color-coded accordingly. Plot (c): Relative energy in the electromagnetic difference field between the superposition and the rigorous numerically calculated fields, to be compared with Fig \ref{fig:udiff} (b,c). } }
\end{figure}

\begin{figure}[bt]
\centering
\includegraphics[width=\figwidth]{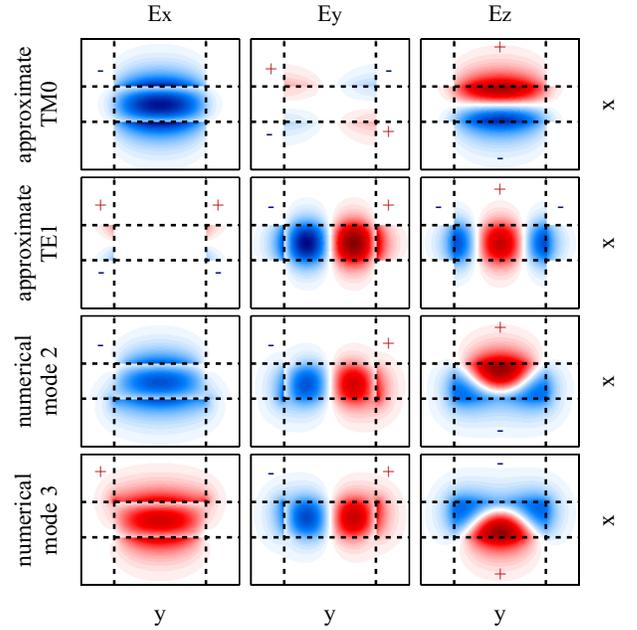} 
\caption{\label{fig:crossingfields} \RevA{Electric fields of different modes in a 833~nm wide by 300~nm high waveguide. The dashed lines separate the different regions (see Fig. 1). Modes $\bm{E}_{\textup{TM0}}$, $\bm{E}_{\textup{TE1}}$, $\bm{E}^N_2$ and $\bm{E}^N_3$ are plotted from top to bottom. Color indicates the field strength, white regions have a low field strength. Regions with positive field strength are indicated with a plus (+) sign and red. Regions with a negative field strength are indicated with a minus sign (-) and blue. } }
\end{figure}

\subsection{\RevA{The avoided crossing phenomenon}}
\label{sec:crossing}
\RevA{
Figures \ref{fig:neff} and \ref{fig:udiff} suggest that something interesting happens at the apparent crossing 
of the effective indices of the TM0-like and the TE1-like modes when the width of the guide is changed. A detailed inspection of the waveguides with widths around 833~nm is presented in Fig.~\ref{fig:crossing}. In plot (a), it can be seen that the numerically computed effective indices of the 2nd and 3rd mode in the waveguide (counted from high to low effective index) actually do not cross each other, but show a behavior that is known in quantum mechanics as \textit{avoided crossing} \cite{refs:landauqm}. We investigate the modes that where found numerically, $\bm{E}^N_2$ and $\bm{E}^N_3$, in terms of the analytically computed approximate modes \RevB{of the waveguide}. 
}

\RevA{
We will verify that the actual modes $\bm{E}^N_2$ and $\bm{E}^N_3$ can in good approximation be written as a superposition of the TM0-like, $\bm{E}_{\textup{TM0}}$, and TE1-like, $\bm{E}_{\textup{TE1}}$ modes that where calculated using the approximate theory (\textmn{amplitude optimization method}), i.e. 
\begin{equation} \label{eq:IVBsuperp}
\bm{E}_i^{\RevB{N}} \RevB{\approx} a \bm{E}_{\textup{TM0}} + b \bm{E}_{\textup{TE1}}, 
\end{equation}
for some real $a$ and $b$, and $i$~=~2~or~3. The phase of mode $\bm{E}_i$ is chosen such that coefficient $b$ is positive. The coefficient $a$ of the TM0-like mode can be either positive or negative. The approximate calculated modes $\bm{E}_{\textup{TM0}}$ and $\bm{E}_{\textup{TE1}}$, are in good approximation orthonormal such that normalization of $\bm{E}_i$ in the norm of Eq. (\ref{eq:TotalFlux}) implies $b = \sqrt{1-a^2}$. 
}

\RevA{  
The coefficient $a$ of the TM0-like mode is optimized such that the difference \RevB{measured using Eq.~(\ref{eq:FieldsDiff}) between the left- and right-hand sides of Eq.~(\ref{eq:IVBsuperp}) is minimum.}
The result is plotted in Fig.~\ref{fig:crossing}(b), 
where it can be seen that mode 2 looks like a TM0-mode at the left of the crossing, 
while it looks like a TE1-like mode on the right-hand-side of the crossing, 
whereas close to the crossing the modes are an equal mixture of $\bm{E}_{\textup{TM0}}$ and $\bm{E}_{\textup{TE1}}$.
In Fig.~\ref{fig:crossing}(c), it can be seen that the error between the superposition $\bm{E}_i$ and the rigorous numerically calculated field $\bm{E}^N_i$ close to the apparent crossing is small and similar to the error that was found away from the crossing (see Fig.~\ref{fig:udiff}). Therefore we may indeed conclude that the field around the crossing can be written as a superposition of modes of the types that are present away from the crossing. Figure \ref{fig:crossingfields} presents the electric fields $\bm{E}_{\textup{TM0}}$, $\bm{E}_{\textup{TE1}}$, $\bm{E}^N_2$ and  $\bm{E}^N_3$ for a 833~nm wide by 300~nm high waveguide, where ${a^2 \approx b^2 \approx 0.5}$. 
}

\RevA{
Using this observation, we will derive a qualitative description of this \textit{avoided crossing}. \RevB{The exact description of propagation of light through waveguides that are invariant in the z-direction can be formulated as an eigenvalue problem with the propagation constant as eigenvalue. In fact, by eliminating the longitudinal components $E_z$ and $H_z$ from Maxwell's equations, one obtains the following eigenvalue problem for the transverse components only }
\RevB{
\begin{equation}\label{eq:betaeve}
\hat{O} 
\begin{pmatrix}
 E_x(x,y) \\ 
 E_y(x,y) \\
 H_x(x,y) \\
 H_y(x,y)   
\end{pmatrix} 
 = \beta_i 
 \begin{pmatrix}
 E_x(x,y) \\ 
 E_y(x,y) \\
 H_x(x,y) \\
 H_y(x,y) 
\end{pmatrix} 
\end{equation}
}%
\RevB{%
with the propagation constant $\beta_i$ as eigenvalue and with $\hat{O}$ a second order partial differential operator with respect to transverse variables $x$ and $y$. We consider forward propagating waves with positive $\beta_i$. The operator $\hat{O}$ is not symmetric, however it can be shown that solutions of Eq.~(\ref{eq:betaeve}) are orthogonal with respect to the bilinear form derived from the power flux \cite{refs:yeh}}
\begin{equation} \label{eq:TotalFlux2}
\braket{i}{j} \equiv \iint \limits_{-\infty}^{\infty} \left( \bm{E}_i \times \bm{H}_j^*  \right) \cdot \bm{\hat{z}} \; d\textup{x}d\textup{y}  = \delta_{ij}
\end{equation}
where the inner product between two solutions \RevB{$i$ and $j$} is defined\RevB{, }$\delta_{ij}$ is the Kronecker delta function, \RevB{and the eigenfields have been normalized}. We \RevB{adopt a \textit{bra-ket}} notation and \RevB{write Eq.~(\ref{eq:betaeve}) as}
\begin{equation}\label{eq:betaeve2}
\hat{O} \ket{i} = \beta_i \ket{i}
\end{equation}
}

\RevA{
We now \RevB{apply} the \RevB{aforementioned} observation that \RevB{in good approximation} the electromagnetic fields of the modes in the waveguide, also around the crossing, can be written as a superposition of the approximate fields, \RevB{hence}
\begin{equation}\label{eq:superpos}
\bm{E_i} \approx a \bm{E}_{\textup{TM0}} + b \bm{E}_{\textup{TE1}}, \;\;\; \textup{or} \;\;\; \ket{i} \approx a \ket{a} + b \ket{b},
\end{equation}
where $\ket{a}$ and $\ket{b}$ represent the TM0-like and TE1-like modes in the waveguide, respectively, while $\ket{i}$ represents the exact solutions of Eq. \RevB{(\ref{eq:betaeve2})}. We only consider the 2nd and 3rd approximate solutions \RevB{here}. \RevB{As will} become clear, only modes with similar propagation constants have to be taken into account around the crossing, whereas the other modes are already accurately calculated \RevB{by} the approximate methods presented in this paper.
Substituting Eq. (\ref{eq:superpos}) in Eq. \RevB{(\ref{eq:betaeve2})} and taking the inner product with $\bra{a}$ gives
\begin{equation}
a \braket{a}{\hat{O}a} + b \braket{a}{\hat{O}b} \approx \RevB{\beta_i} \left(a \braket{a}{a} + b \braket{a}{b} \right).
\end{equation}
If we also take the inner product of Eq. \RevB{(\ref{eq:betaeve2})} with $\bra{b}$ we arrive at the (2x2)-system: 
\begin{equation}
\begin{pmatrix}
 \braket{a}{\hat{O}a} & \braket{a}{\hat{O}b} \\ 
 \braket{b}{\hat{O}a} & \braket{b}{\hat{O}\RevB{b}}  
\end{pmatrix} 
\begin{pmatrix}
 a \\ 
 b  
\end{pmatrix} 
\approx \RevB{\beta_i}
\begin{pmatrix}
 \braket{a}{a} & \braket{a}{b} \\ 
 \braket{b}{a} & \braket{b}{b}  
\end{pmatrix} 
\begin{pmatrix}
 a \\ 
 b  
\end{pmatrix}, 
\end{equation}
or, 
\begin{equation}\label{eq:M1}
M \begin{pmatrix} a \\ b \end{pmatrix} 
\approx \RevB{\beta_i}
\begin{pmatrix} a \\ b \end{pmatrix},
\end{equation}
where $M = $
\small 
\begin{equation} \nonumber
\frac{1}{D} 
\begin{pmatrix}
 \braket{b}{b}\braket{a}{\hat{O}a} - \braket{a}{b}\braket{b}{\hat{O}a} &  \braket{b}{b}\braket{a}{\hat{O}b} - \braket{a}{b}\braket{b}{\hat{O}b}\\
-\braket{b}{a}\braket{a}{\hat{O}a} + \braket{a}{a}\braket{b}{\hat{O}a} & -\braket{b}{a}\braket{a}{\hat{O}b} + \braket{a}{a}\braket{b}{\hat{O}b}  
\end{pmatrix} 
\end{equation}
\normalsize
and %
\begin{equation} \nonumber
D = \braket{a}{a}\braket{b}{b} - \braket{a}{b}\braket{b}{a}.
\end{equation}
}%
\RevA{
The modes that we found in our approximate analysis are almost orthonormal, so $\braket{a}{a}$ and $\braket{b}{b}$ are approximately unity and $\braket{a}{b}$ and $\braket{b}{a}$ are approximately zero. Away from the crossing, we found that the approximate solutions $\ket{a}$ and $\ket{b}$ obey relation Eq. \RevB{(\ref{eq:betaeve2})} so that $\braket{a}{\hat{O}a} \approx \beta_a$, $\braket{b}{\hat{O}b} \approx \beta_b$, while $\braket{a}{\hat{O}b}$ and $\braket{b}{\hat{O}\RevB{a}}$ are small. This allows us to write the relation Eq. (\ref{eq:M1}) as 
\begin{equation}\label{eq:M2}
\begin{pmatrix}
\beta_a + \delta_a & \delta_{ab} \\
\delta_{ba} & \beta_b + \delta_b
\end{pmatrix} 
\begin{pmatrix} a \\ b \end{pmatrix} 
\approx \RevB{ \beta_i }
\begin{pmatrix} a \\ b \end{pmatrix},
\end{equation}
where $\delta_a$, $\delta_b$, $\delta_{ab}$, and $\delta_{ba}$ are quantities that are much smaller than the $\beta$'s. 
This system has the eigenvalues \cite{refs:landauqm}
\begin{equation}
\beta_{2,3} = \frac{\beta_a' + \beta_b'}{2} \pm \frac{\sqrt{(\beta_a' - \beta_b')^2 + 4 \delta_{ab} \delta_{ba}}}{2},
\end{equation}
with corresponding eigenvectors $\bm{v}_{2,3}$ \RevB{(not normalized)}
\small
\begin{equation}
\begin{pmatrix}
2 \RevB{\delta_{ab}}\\ 
-\beta_a' + \beta_b' \pm \sqrt{(\beta_a'- \beta_b')^2 + 4 \delta_{ab} \delta_{ba}}
\end{pmatrix},
\end{equation}
\normalsize
where ${\beta_a' \equiv \beta_a + \delta_a}$ and ${\beta_b' \equiv \beta_b + \delta_b}$.
The two propagation constants are closest when ${\beta_a' = \beta_b'}$ but are always \RevB{separated} by a minimum distance ${\RevB{4} \sqrt{\delta_{ab} \delta_{ba}}}$, so that they never intersect. For small ${\RevB{\delta_a, \delta_b, \delta_{ab}, \delta_{ba}} \ll |\beta_a - \beta_b|}$, we find the eigenvector for ${\beta_a > \beta_b}$ to be ${\bm{v}_2 \approx \RevB{(}1, 0\RevB{)}}$ and ${\bm{v}_3 \approx \RevB{(}0, 1\RevB{)}}$. 
The upper propagation constant, $\beta_2$, has a TM0-like mode in this limit, while the lower propagation constant, $\beta_3$, has a TE1-like mode. For ${\beta_b > \beta_a}$ we find ${\bm{v}_2 \approx \RevB{(}0, 1\RevB{)}}$ and ${\bm{v}_3 \approx \RevB{(}1, 0\RevB{)}}$, so that the upper propagation constant now has a TE1-like mode while the lower propagation constant has a TM0-like mode. An interesting \RevB{case occurs when} ${\beta_a' = \beta_b'}$ \RevB{and $\delta_{ab} = \delta_{ba}$. Then} the \RevB{normalized} eigenvectors of this system are 
${\bm{v}_2 = \frac{1}{\sqrt{2}} \cdot \RevB{(}1, 1\RevB{)}}$ and 
${\bm{v}_3 = \frac{1}{\sqrt{2}} \cdot \RevB{(}1, -1\RevB{)}}$, i.e. they are either an addition or subtraction of the eigenvectors far from the crossing. 
}

\RevA{
This simple description of the \textit{avoided crossing} agrees with the observations of the numerically computed modal profiles as presented in Figs. \ref{fig:crossing} and \ref{fig:crossingfields}.
}

\begin{figure}[thp]
\centering
\includegraphics[width=\figwidth]{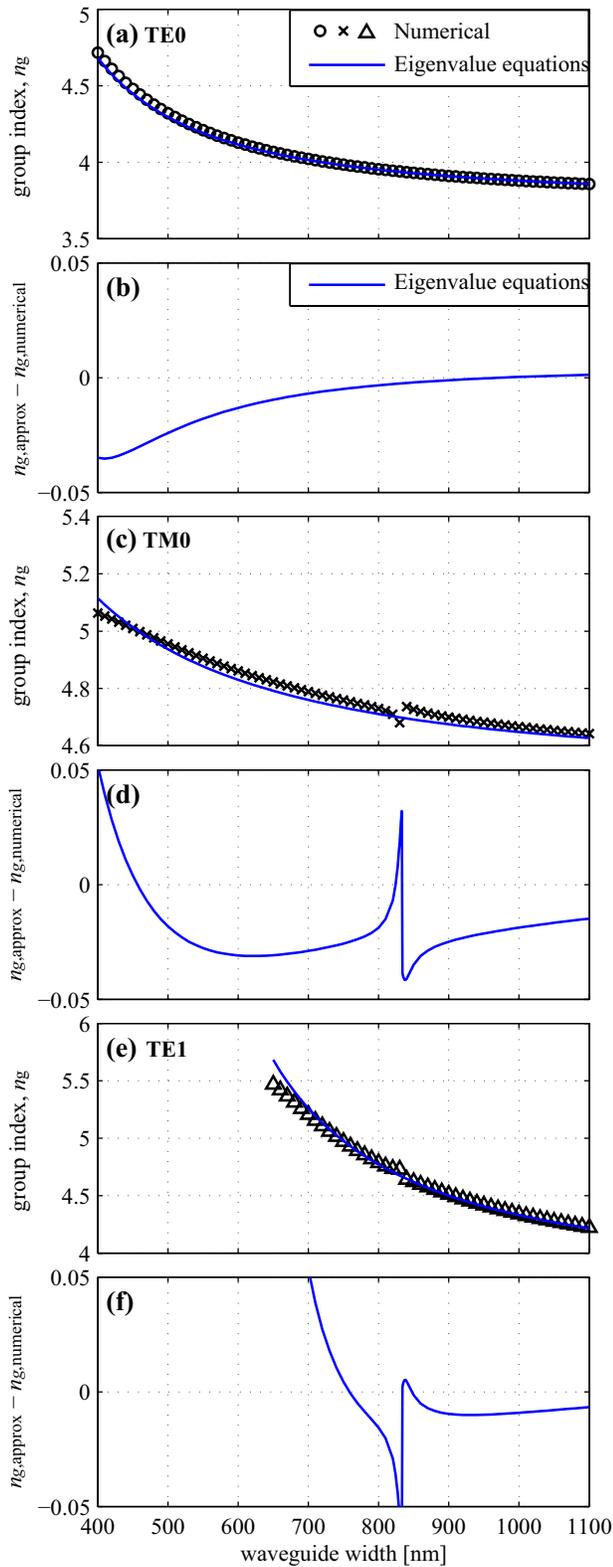} 
\caption{\label{fig:ng} Effective group indices. Plots (a), (c), (e) present the numerical result, as well as the result of the extension of Marcatili's method (eigenvalue equation). Plots (b), (d), (f) show the difference between the two methods. }
\end{figure}

\begin{figure}[tbhp]
\centering
\includegraphics[width=\figwidth]{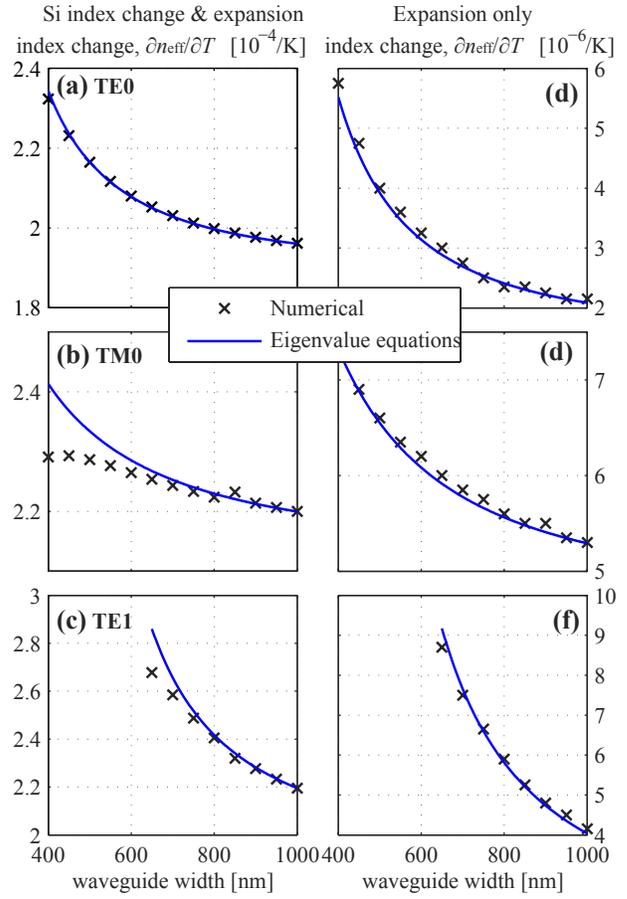} 
\caption{\label{fig:tshift} Shift in effective refractive index due to temperature variation. (Left plots) Shift in effective refractive index due to change in silicon refractive index (thermo-optic effect) as well as expansion of the waveguide (thermal expansion) is depicted. (Right plots) Only the geometry of the waveguide is varied (thermal expansion), with the silicon refractive being constant. The result of the extension of Marcatili's method, and the result of the numerical FMM method are presented.}
\end{figure}

\begin{figure}[tbhp]
\centering
\includegraphics[width=\figwidth]{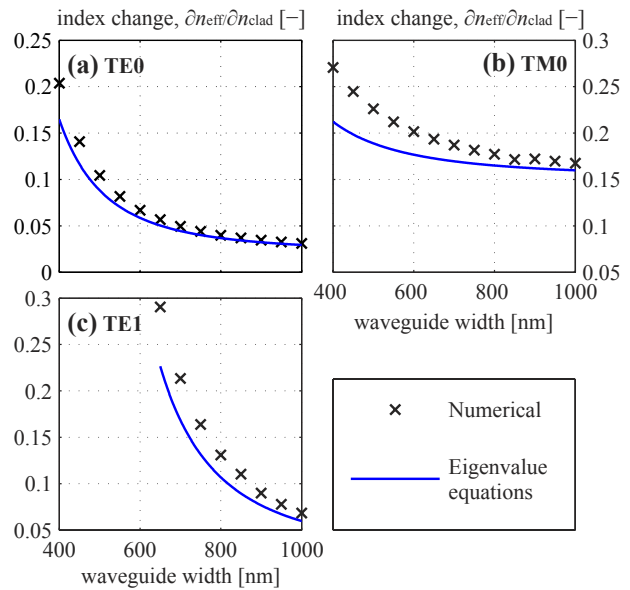} 
\caption{\label{fig:nshift} Shift in effective index due to change in cladding refractive index. The unit is refractive index change per cladding material index change [-/-]. The initial cladding is water ($n=1.31$).}
\end{figure}

\subsection{Novel applications of the eigenvalue equation of the propagation constant}
We applied the analytical eigenvalue equations for the effective index of the modes in rectangular waveguide to explicitly calculate of the effective group index (Sec. \ref{sec:modaldispersion}), as well as the linearized influence of an external effect (Sec. \ref{sec:rectexternaleffect}). In this section, we compare those analytical results with rigorous numerical simulations. For numerical calculation of the linearized influence of an external effect on the effective index of the modes in the waveguide, we use the following procedure: first, we calculated the effective refractive index ($n_{\textup{eff}}$) for different values of the external effect ($\chi$) that are around the starting value ($\chi_0$), then, a linear fit is used to obtain the $\partial n_{\textup{eff}}/\partial \chi$. 

\subsubsection{Effective group index}
The method presented in Sec. \ref{sec:modaldispersion} is investigated in this section. Equations (\ref{eq:groupindex}), (\ref{eq:dkxdk0b}) and (\ref{eq:dkydk0b}) are used to calculate the effective group index of the modes in the waveguides. The silicon dispersion is taken into account, with $\partial n_1 / \partial k_0 =  3.147 \cdot 10^8$ m$^{-1}$ \cite{refs:fimmwavematerial}. These results are compared with the group index that was numerically calculated using the FMM method as implemented in FimmWave. Results are presented in Fig.~\ref{fig:ng} where can be seen that the error remains below 4\%.

\subsubsection{Temperature-induced effective index change}
\label{sec:tnumerical}
To rigorously compute the linearized influence of temperature variations on the effective index of the waveguide, the temperature in the simulation is varied from 20 $^\circ$C to 28 $^\circ$C in steps of 2 $^\circ$C. The influence $\partial n_{\textup{eff}}/\partial T$ is compared with the analytically calculated influence in Fig.~8. Two effects are taken into account: the thermo-optic effect using $\partial n_1 / \partial \chi = 1.83 \cdot 10^{-4}$ $^\circ$C$^{-1}$, and the thermal expansion using $\partial b / \partial \chi = \alpha b$ m/$^oC$ and $\partial d / \partial \chi = \alpha d$ m/$^\circ$C, with linear thermal expansion coefficient $\alpha = 2.6 \cdot 10^{-6}$ $^\circ$C$^{-1}$ \cite{refs:okada84,refs:fimmwavematerial}. The effects are investigated simultaneous (net effect) and the effect of the expansion only is also shown separately. This separate investigation does, of course, not simulate a physical measurement but it gives a good test case for the method. From Fig.~\ref{fig:tshift}, it can be concluded that the effects are well estimated with a relative error below 7\%. The error is much smaller for modes that are less confined, and the effect of the geometrical change is better estimated than the effect of the silicon index change.

\subsubsection{Cladding-index induced effective index change}
In Sec. \ref{sec:externalcladding}, we described the linearized influence of a change in refractive index of the cladding of the waveguide. This is, for example, used in evanescent field sensors, where a liquid flows over the waveguide surface. 
For example, aqueous solutions with different concentrations of sucrose or sodium chlorine are used in Refs.  \cite{refs:densmore06} and \cite{refs:devos07}, respectively.
The rigorous numerical results are obtained by changing the cladding refractive index from 1.310 (water) to 1.320 in steps of 0.002. From this, $\partial n_{\textup{eff}}/ \partial n_{\textup{clad}}$ is extracted. Results are shown in Fig.~\ref{fig:nshift}, where it can be seen that relative errors are up to 23\%. These errors rapidly decrease for wider waveguides, i.e. for less-confined modes.

\section{Conclusion}
\label{sec:conclusion}
We derived a model for high-index-contrast rectangular dielectric waveguides, based on an \textit{ansatz} that describes the fields in the core of the waveguide as standing waves. As in Marcatili's work, we also arrive at eigenvalue equations for the spatial frequencies of the standing waves, i.e. $k_x$ and $k_y$, that are identical to the eigenvalue equations of slab waveguides. In order to obtain $k_x$, we use the limit that the waveguide extends to infinity in the y-direction, and order to obtain $k_y$, we use the limit that the waveguide extends to infinity in the x-direction. The effective refractive index of the mode in the waveguide, which directly follows from $k_x$ and $k_y$, agrees excellently with rigorous numerical mode solvers (relative error $<$ 2\%). In comparison with Marcatili's original work, our novel choice of electromagnetic field amplitudes in the \textit{ansatz} removed the discontinuity of the dominant electromagnetic field components, which was severe for high-index-contrast waveguides. This improvement led to better agreement with numerical simulations, with a relative energy of the difference field below 3\%, \RevA{except when the effective indices of two modes in the waveguide are similar}. 
We quantified the error that arises from the discontinuity of the fields in terms of energy density. With this quantification, we accurately predicted the trend in the error of the different analytical methods. In addition, this quantification allowed us to optimize the amplitudes in the \textit{ansatz} such that minimal mismatch is achieved. Our  \textmn{amplitude optimization method} accurately describes both TE-like as well as TM-like modes in the rectangular waveguides \RevA{(except when two modes have similar effective indices)}, and the error of this method with respect to the numerical results is very close to the fundamental error in the \textit{ansatz}. 

Next to this, we derived explicit expressions for the effective group index of the waveguide, taking dispersion into account. The error in comparison with rigorous numerical methods was below 4\%. Also, explicit expressions are derived for the linearized influence of an external effect. We predicted the influence of temperature, as well as the influence of a change of the refractive index of the cladding of the waveguide. 

We applied our method to interpret the results of a rigorous numerical mode solver, \RevA{and showed that the modes in the waveguide show \textit{avoided crossing} behavior when the effective indices of a TE-like and a TM-like mode are similar}. A quick analysis of waveguides can be performed with this method, while the results of this method can be used as starting point for rigorous mode solvers. Marcatili's method has long been used for educational purposes, as is indicated by its frequent occurrence in textbooks. With our work, the development of intuitive analytical methods now follows the technological developments to high-index-contrast waveguides.

\appendices

\begin{figure*}[!t]
\normalsize
\setcounter{MYtempeqncnt}{\value{equation}}
\setcounter{equation}{76}
\begin{align}
\label{eq:dkxdk0b}
\frac{\partial k_x}{\partial k_0} = & \frac{
k_x k_0
\left( 
\alpha_2 (n_1^2 - n_2^2) + \alpha_3 (n_1^2 - n_3^2) + \left(\alpha_2 + \alpha_3 \right) n_1 k_0 \dd{n_1}{k_0}
\right )
+ \left( \frac{\gamma_2}{n_2^2} + \frac{\gamma_3}{n_3^2} \right) 
\left( \frac{4 k_x^3}{n_1^5} - \frac{2 k_x}{n_1} \left( \frac{k_x^2}{n_1^4} - \frac{\gamma_2 \gamma_3}{n_2^2 n_3^2} \right) \right) \dd{n_1}{k_0}
}{
\left( \frac{\gamma_2}{n_2^2} + \frac{\gamma_3}{n_3^2} \right)
\left( \frac{k_x^2}{n_1^4} + \frac{\gamma_2 \gamma_3}{n_2^2 n_3^2} \right)
+ k_x^2 \left( \alpha_2 + \alpha_3 \right) 
+ n_1^2 d \left( \frac{k_x^2}{n_1^4} - \frac{\gamma_2 \gamma_3}{n_2^2 n_3^2} \right)^2 \sec^2[k_x d]
} \\
\label{eq:dkydk0b}
\frac{\partial k_y}{\partial k_0} = & \frac{
k_y k_0 \left( \alpha_4 (n_1^2 - n_4^2) +
							 \alpha_5 (n_1^2 - n_5^2) +
							 \left(\alpha_4 + \alpha_5 \right) n_1 k_0 \dd{n_1}{k_0}
				\right )
}{
(\gamma_4 + \gamma_5)(k_y^2 + \gamma_4 \gamma_5)
+ k_y^2 \left( \alpha_4 + \alpha_5 \right) 
+ b (k_y^2 - \gamma_4 \gamma_5)^2 \sec^2[k_y b]
}
\end{align}
\setcounter{equation}{80}
\begin{align}
\label{eq:dkxdXb}
\begin{split}
\frac{\partial k_x}{\partial \chi} = &
k_x \left\{
\left( (\alpha_2 + \alpha_3) k_0^2 n_1 + \alpha_a \left( \frac{4 k_x^2}{n_1^5} - \frac{2 \alpha_b^2}{n_1} \right) \right) \dd{n_1}{\chi} 
- \left( \alpha_2 k_0^2 n_2 + 2 \frac{\gamma_2}{n_2^3} \left( \alpha_b^2 + \alpha_a \frac{\gamma_3}{n_3^2} \right) \right) \dd{n_2}{\chi}
\right. \\ 
- & \left. \left( \alpha_3 k_0^2 n_3 + 2 \frac{\gamma_3}{n_3^3} \left( \alpha_b^2 + \alpha_a \frac{\gamma_2}{n_2^2} \right) \right) \dd{n_3}{\chi} 
- \alpha_b^4 n_1^2 k_x \sec^2[k_x d] \dd{d}{\chi} 
\right\}
\\
& \cdot \left\{ 
\alpha_a
\left( \frac{k_x^2}{n_1^4} + \frac{\gamma_2 \gamma_3}{n_2^2 n_3^2} \right)
+ k_x^2 \left( \alpha_2 + \alpha_3 \right) 
+ n_1^2 \alpha_b^4 d \sec^2[k_x d]
\right\}^{-1}
\end{split}
\\
\label{eq:dkydXb}
\frac{\partial k_y}{\partial \chi} = & \frac{
k_y k_0^2 \left( (\alpha_4 + \alpha_5) n_1 \dd{n_1}{\chi} 
	- \alpha_4 n_4 \dd{n_4}{\chi} - \alpha_5 n_5 \dd{n_5}{\chi} 
	\right )
	- k_y (k_y^2 - \gamma_4 \gamma_5)^2 \sec^2[k_y b] \dd{b}{\chi}
}{
(\gamma_4 + \gamma_5)(k_y^2 + \gamma_4 \gamma_5)
+ k_y^2 \left( \alpha_4 + \alpha_5 \right) 
+ b (k_y^2 - \gamma_4 \gamma_5)^2 \sec^2[k_y b]
},
\end{align}
\setcounter{equation}{\value{MYtempeqncnt}}
\hrulefill
\vspace*{4pt}
\end{figure*}

\section{Explicit equation for modal dispersion}
\label{sec:Amodaldispersion}

In Sec. \ref{sec:modaldispersion}, partial derivatives $\partial k_x / \partial k_0$ and $\partial k_x / \partial k_0$ were given by Eqs. (\ref{eq:dkxdk0}) and (\ref{eq:dkydk0}). In this section, we give $\partial k_x/\partial k_0$ and $\partial k_y/\partial k_0$ for the case when only the core material is dispersive, i.e. $n_1(k_0)$ and where the other refractive indices do not depend on the frequency, thus also not on $k_0$. We define:
\begin{align}
\alpha_2 \equiv & \left( \frac{k_x^2}{n_1^4}+\frac{\gamma_3^2}{n_3^4} \right) \frac{1}{n_2^2 \gamma_2}, \label{eq:alpha2}\\
\alpha_3 \equiv & \left( \frac{k_x^2}{n_1^4}+\frac{\gamma_2^2}{n_2^4} \right) \frac{1}{n_3^2 \gamma_3},\\
\alpha_4 	\equiv & \frac{k_y^2+\gamma_5^2}{\gamma_4}, \\
\alpha_5 	\equiv & \frac{k_y^2+\gamma_4^2}{\gamma_5}. \label{eq:alpha5}
\end{align}%
And find that Eqs. (\ref{eq:dkxdk0}) and (\ref{eq:dkydk0}) are explicitly given by Eqs. (\ref{eq:dkxdk0b}) and (\ref{eq:dkydk0b}), which are shown on the next page.
\addtocounter{equation}{2}

\section{Explicit equation for external effect}
\label{sec:Aexternaleffect}
In this appendix, we give the explicit form of Eqs. (\ref{eq:dkxdX}) and (\ref{eq:dkydX}). Let:
\begin{align}
\alpha_a \equiv & \frac{\gamma_2}{n_2^2} + \frac{\gamma_3}{n_3^2},  \\
\alpha_b^2 \equiv & \frac{k_x^2}{n_1^4} - \frac{\gamma_2 \gamma_3}{n_2^2 n_3^2}.
\end{align}
Then Eqs. (\ref{eq:dkxdX}) and (\ref{eq:dkydX}) are identical to Eqs. (\ref{eq:dkxdXb}) and (\ref{eq:dkydXb}), see next page, where $\alpha_2$-$\alpha_5$ are defined in Eqs. (\ref{eq:alpha2}) - (\ref{eq:alpha5}).
\addtocounter{equation}{2}

\section*{Acknowledgment}
The authors would like to thank dr. \mbox{Omar} \mbox{El Gawhary}, dr. \mbox{Jose} \mbox{Pozo}, \RevA{mr. \mbox{Vincent} \mbox{Brulis}, dr. Jos \mbox{Thijssen}, dr. Ad \mbox{Verbruggen}, and ir. \mbox{Kevin} van \mbox{Hoogdalem}} for fruitful discussions. 

\ifCLASSOPTIONcaptionsoff
  \newpage
\fi


\begin{thebibliography}{10}
\providecommand{\url}[1]{#1}
\csname url@samestyle\endcsname
\providecommand{\newblock}{\relax}
\providecommand{\bibinfo}[2]{#2}
\providecommand{\BIBentrySTDinterwordspacing}{\spaceskip=0pt\relax}
\providecommand{\BIBentryALTinterwordstretchfactor}{4}
\providecommand{\BIBentryALTinterwordspacing}{\spaceskip=\fontdimen2\font plus
\BIBentryALTinterwordstretchfactor\fontdimen3\font minus
  \fontdimen4\font\relax}
\providecommand{\BIBforeignlanguage}[2]{{%
\expandafter\ifx\csname l@#1\endcsname\relax
\typeout{** WARNING: IEEEtran.bst: No hyphenation pattern has been}%
\typeout{** loaded for the language `#1'. Using the pattern for}%
\typeout{** the default language instead.}%
\else
\language=\csname l@#1\endcsname
\fi
#2}}
\providecommand{\BIBdecl}{\relax}
\BIBdecl

\bibitem{refs:Marcatili69}
E.~Marcatili, ``Dielectric rectangular waveguide and directional coupler for
  integrated optics,'' \emph{The Bell System Technical Journal}, vol.~48, pp.
  2071--2121, Mar. 1969.

\bibitem{refs:marcuse}
D.~Marcuse, \emph{Theory of dielectrically optical waveguides}.\hskip 1em plus
  0.5em minus 0.4em\relax San Diego: Academic Press Inc, 1991.

\bibitem{refs:yeh}
C.~Yeh and F.~I. Shimabukuro, \emph{The Essence of Dielectric
  Waveguides}.\hskip 1em plus 0.5em minus 0.4em\relax New York: Springer
  Science+Business Media, LCC, 2008.

\bibitem{refs:pollock}
C.~Pollock and M.~Lipson, \emph{Integrated Photonics}.\hskip 1em plus 0.5em
  minus 0.4em\relax Boston: Kluwer Academic Publishers, 2003.

\bibitem{refs:hunsperger2002}
R.~G. Hunsperger, \emph{Integrated optics: theory and technology}.\hskip 1em
  plus 0.5em minus 0.4em\relax Berlin: Springer-Verlag, 2002.

\bibitem{refs:dumon08}
P.~Dumon, W.~Bogaerts, A.~Tchelnokov, J.-M. Fedeli, and R.~Baets, ``Silicon
  nanophotonics,'' \emph{Future Fab International}, vol.~25, pp. 29--36, Apr.
  2008.

\bibitem{refs:yariv00}
A.~Yariv, ``Universal relations for coupling of optical power between
  microresonators and dielectric waveguides,'' \emph{Electronics Letters},
  vol.~36, no.~4, pp. 321 --322, feb 2000.

\bibitem{refs:smit88}
M.~Smit, ``New focusing and dispersive planar component based on an optical
  phased array,'' \emph{Electronics Letters}, vol.~24, no.~7, pp. 385 --386,
  mar 1988.

\bibitem{refs:goell69}
J.~Goell, ``A circular-harmonic computer analysis of rectangular dielectric
  waveguides,'' \emph{The Bell System Technical Journal}, vol.~48, pp.
  2133--2160, 1969.

\bibitem{refs:subdo94}
A.~S. Sudbo, ``Improved formulation of the film mode matching method for mode
  field calculations in delectric waveguides,'' \emph{Pure and Applied Optics:
  Journal of the European Optical Society Part A}, vol.~3, pp. 381--388, 1994.

\bibitem{refs:hammer07}
O.~Ivanova, M.~Hammer, R.~Stoffer, and E.~van Groesen, ``A variational mode
  expansion mode solver,'' \emph{Optical and Quantum Electronics}, vol.~39, pp.
  849--864, 2007.

\bibitem{refs:fimmwave}
``Fimmwave - numerical waveguide mode solver,'' Photon Design Ltd, Oxford,
  United Kingdom, Jan. 2011.

\bibitem{refs:melloni09}
A.~Melloni, D.~Roncelli, F.~Morichetti, A.~Canciamilla, and A.~Bakker,
  ``Statistical design in integrated optics,'' in \emph{CLEO/Europe and EQEC
  2009 Conference Digest}.\hskip 1em plus 0.5em minus 0.4em\relax Optical
  Society of America, 2009, p. JSI1\_4.

\bibitem{refs:densmore06}
A.~Densmore, D.-X. Xu, P.~Waldron, S.~Janz, P.~Cheben, J.~Lapointe, A.~Delage,
  B.~Lamontagne, J.~Schmid, and E.~Post, ``A silicon-on-insulator photonic wire
  based evanescent field sensor,'' \emph{Photonics Technology Letters, IEEE},
  vol.~18, no.~23, pp. 2520 --2522, dec.1, 2006.

\bibitem{refs:devos07}
K.~D. Vos, I.~Bartolozzi, E.~Schacht, P.~Bienstman, and R.~Baets,
  ``Silicon-on-insulator microring resonator forsensitive and label-free
  biosensing,'' \emph{Opt. Express}, vol.~15, no.~12, pp. 7610--7615, Jun 2007.

\bibitem{refs:fimmwavematerial}
``Material database and material models,'' distributed with FimmWave software
  package, Photon Design Ltd, Oxford, United Kingdom, Jan. 2011.

\bibitem{refs:matlab}
``Matlab - the language of technical computing,'' The MathWorks, Inc., Natick,
  Massachusetts, USA, 2010.

\bibitem{refs:okada84}
Y.~Okada and Y.~Tokumaru, ``Precise determination of lattice parameter and
  thermal expansion coefficient of silicon between 300 and 1500 k,''
  \emph{Journal of Applied Physics}, vol.~56, no.~2, pp. 314--320, 1984.

\bibitem{refs:landauqm}
L.~Landau and E.~Lifshitz, \emph{Quantum Mechanics}, 3rd~ed.\hskip 1em plus
  0.5em minus 0.4em\relax Pergamon Press Ltd., 1977, sec.~79.

\end{thebibliography}
\end{document}